\documentclass[english,a4paper,aps,pre,reprint,amsmath,amssymb,superscriptaddress,nofootinbib]{revtex4-2}
\usepackage{preamble}
\graphicspath{{figures/}}

\begin{document}

\title{Universal limiting behaviour of reaction-diffusion systems with conservation laws}
\date{\today}
\author{Joshua F. Robinson}
\affiliation{The Hartree Centre, STFC Daresbury Laboratory, Warrington, WA4 4AD, United Kingdom}
\affiliation{H.\ H.\ Wills Physics Laboratory, University of Bristol, Bristol BS8 1TL, United Kingdom}
\email{joshua.robinson@stfc.ac.uk}
\author{Thomas Machon}
\affiliation{H.\ H.\ Wills Physics Laboratory, University of Bristol, Bristol BS8 1TL, United Kingdom}
\author{Thomas Speck}
\affiliation{Institute for Theoretical Physics IV, University of Stuttgart, 70569 Stuttgart, Germany}

\begin{abstract}
  Making sense of complex inhomogeneous systems composed of many interacting species is a grand challenge that pervades basically all natural sciences.
  Phase separation and pattern formation in reaction-diffusion systems have been largely studied as two separate paradigms.
  Here we show that in reaction-diffusion systems composed of many species, the presence of a conservation law constrains the evolution of the conserved quantity to be governed by a Cahn-Hilliard-like equation.
  This establishes a direct link with the paradigm of coexistence and recent ``active'' field theories.
  Hence, even for complex many-species systems a dramatically simplified but accurate description emerges over coarse spatio-temporal scales.
  Using the nullcline (the line of homogeneous steady states) as the central motif, we develop a geometrical framework which endows chemical space with a basis and suitable coordinates.
  This framework allows us to capture and understand the effect of eliminating fast non-conserved degrees of freedom, and to explicitly construct coefficients of the coarse field theory.
  We expect that the theory we develop here will be particularly relevant to advance our understanding of biomolecular condensates.
\end{abstract}

\maketitle

\section{Introduction}

Life is rich with beautiful patterns, from the organisation of cells in growing tissues and bacterial colonies to the stripes of the humble zebra.
An intriguing recent example is the role of liquid-like protein condensates in cell biology.
While compartmentalisation can be realised through membranes, there is now ample evidence that nature also utilises the physics of phase transitions to recruit or sequester proteins at certain locations in the form of membrane-less organelles~\cite{hyman2014,banani2017,weber2019,falahati2019}.
These \emph{condensates} exchange particles with their surroundings yet remain stable (in finite numbers) over the lifetime of the cell \cite{cho2018}. \emph{In vivo,} they do not grow by Ostwald ripening, precluding explanation in terms of an equilibrium (oil-in-water) mechanism for droplet formation.
Despite this, oil-in-water still is very much the paradigm prevailing in the biological literature \cite{hyman2014,banani2017}, although chemical activity has been proposed as a mechanism to stabilise condensates at fixed sizes \cite{zwicker2017,wurtz2018,leeJPD2019,bressloff2020,bauermann2022,zwicker2022,bergmann2023}.
Establishing a deeper theoretical understanding within existing approaches is challenged by the shear number of (molecular) components and calls for novel theoretical approaches that take into account the non-equilibrium nature of living matter.

\citeauthor{turing1952}'s seminal work established that patterns can spontaneously emerge from coupling chemical reactions to diffusion \cite{turing1952}.
A reaction-diffusion system describes the evolution of an $m$-component chemical field $\vec{\rho} = (\rho^1, \dots, \rho^m)^\top \in \chemicalspace \cong \mathbb{R}^m$ in $d$ spatial dimensions.
These components represent the concentrations of ``species'': bacteria, small molecules that react, enzymes and other biomolecules that undergo transitions between conformations (that modify how they interact), etc.; and must therefore be non-negative $\rho^i \ge 0$.
The system evolves according to
\begin{equation}\label{eq:evolution}
  \partial_t \vrho = \vRflux + \LD \nabla^2 \vrho\,,
\end{equation}
where $\vRflux = \vRflux(\vrho)$ is the chemical flux (\ie $R^i$ is the net change of species $i$ per time) and $\LD$ is a symmetric and positive-definite matrix of transport coefficients that may include cross-diffusion.
Specific examples like the Gray-Scott model~\cite{gray1983,gray1984} and the complex Ginzburg-Landau equation~\cite{aranson2002} only describe a few (effective) components lacking conservation laws. This also holds for models in the same spirit such as Swift-Hohenberg~\cite{swift1977} that contain higher spatial derivatives.
More generally, systems with conserved and non-conserved terms that interact via forces also exhibit patterns \cite{glotzer1995}.
Other chemical systems can form patterns without strictly being via the Turing mechanism, \eg in response to externally-imposed chemical gradients \cite{ouyang1989}.

The Turing mechanism has been demonstrated in bespoke \textit{in vitro} experiments \cite{castets1990}, but its evidence in real biological systems such as seashells or skin pigmentation is weak \cite{kondo2010,kondo2021}.
That being said, its simplicity makes it ideally suited as a starting point to theoretical study \eg cell polarisation models~\cite{otsuji2007,altschuler2008,goryachev2008,mori2008,jilkine2011,edelstein-keshet2013,trong2014,seirinlee2015,chiou2018,bergmann2018,bergmann2019,miller2023}.
Although he did not articulate it in thermodynamic terms, \citeauthor{turing1952} essentially discovered a non-equilibrium mechanism for pattern formation that relies on dissipative driving forces.
There is a direct line of continuity from \citeauthor{turing1952}'s work through foundational results in non-equilibrium thermodynamics \cite{prigogine1967,prigogine1968} which continues to influence the modern day fields of stochastic thermodynamics \cite{seifert2012} and active matter \cite{ramaswamy2010,marchetti2013,bechinger2016}.

The number of components in any real biological system will be very large with $m \gg 1$, and so we seek a reduced description in terms of a more manageable subset of \emph{order parameters}.
In biological systems, we anticipate the numbers of some macromolecules to be effectively conserved on the timescale of structure formation. We say a quantity $\phi$ is conserved if its evolution follows the continuity equation $\partial_t\phi+\nabla\cdot\vec J=0$ with current $\vec J$.
We expect conserved quantities to be good candidates for order parameters since any change has to be due to exchange, which makes their (coarse-grained) evolution slow compared to non-conserved components.
Timescale separation is essential: e.g., condensates in the cell nucleus form on a timescale less than $\order{\SI{1}{\second}}$ \cite{cho2018}.
By contrast, even a relatively small protein comprising 100 amino acids will typically take $\order{\SI{10}{\second}}$ to synthesise \cite{milo2016}, and potentially 1-2 orders of magnitude longer to mature into its functional state \cite{milo2016,balleza2018}.
Many condensates involve proteins that are an order of magnitude larger.
For example, the protein BRD4 comprises $\sim 750$ amino acids \cite{hanNSMB2020}, implying a synthesis time around $\order{\SI{100}{\second}}$, and plays a central role in the formation of transcriptional condensates which have lifetimes typically two orders of magnitude smaller \cite{cho2018,hanNSMB2020}.
As such, we expect the total number of proteins involved in self-organisation of condensates to be essentially conserved over experimental timescales relevant to determining phase behaviour.
Simpler chemical species, such as ATP/ADP, will fluctuate in number on these same timescales, but will be constrained by the (effectively conserved) number of proteins which regulate biological processes.

Our heuristic argument above suggests that approximate conservation laws are at least plausible in biological systems.
If the reaction-diffusion equation \eqref{eq:evolution} respects such a conservation law, the question remains as to how non-equilibrium constraints are imposed.
In purely reactive systems (without diffusion), these are typically imposed by making the system ``open'' through introducing distinct and kinetically isolated chemical reservoirs that are out of equilibrium (see \Refcite{rao2016} and references therein).
Chemostat exchanges with each reservoir \ce{A_i <=> $\emptyset$} introduce terms inside the chemical flux $\vRflux$.
A common approach is to assume these happen infinitely fast, fixing the concentrations $[A_i]$.
An example of this is the Willamowski-R\"ossler model \cite{willamowski1980}.
However, in reaction-diffusion there is an additional mechanism for non-equilibrium behaviour: a mismatch between the shape of $\vRflux$ and the \emph{ideal} free energy implied by the diffusive term.
The flux $\vRflux$ would have to be trivially linear in $\vrho$ for consistency with the latter term.
Consequently, any nonlinear form for $\vRflux$ imposes a non-equilibrium constraint on the thermal and chemical reservoirs.
As this is already non-trivial, we will focus on the general case of nonlinear $\vRflux$ without getting into details of the underlying chemical reservoirs.
We are therefore agnostic to whether the flux is consistent with equilibrium dynamics (a single chemical reservoir) or non-equilibrium (multiple).

Pattern formation in the presence of such conservation laws has recently been addressed by Frey and coworkers \cite{halatek2018,brauns2020,brauns2021}, providing a foundational framework for the phase-space features underlying patterns far from equilibrium.
They mostly focus on archetypal two-component systems.
Here we build on their geometric insights and construct explicit evolution equations that bridge reaction-diffusion systems (with arbitrarily many-species) and recent field theories for active matter.
This approach relies on timescale separation, becoming asymptotically exact in the limit where relaxation of the conserved quantity occurs on the slowest timescale.
We will assume this regime.
Proving this regime bounds trajectories would require more sophisticated techniques - see \eg \Refcite{mallet-paret1988} and references therein.
In particular, the conservation law implies the existence of a ``slow manifold''\footnote{A slow manifold is a type of invariant manifold: a global attractor of trajectories. A slow manifold is characterised by also being marginally stable about the steady-state. A genuine conservation law implies marginal stability everywhere, so guarantees the existence of a slow manifold in chemical space for the homogeneous kinetics.} in the chemical flux (for the homogeneous kinetics $\partial_t \vrho = \vRflux$).
Slow manifolds have been studied in the context of \emph{homogeneous} enzyme kinetics \cite{roussel2001}, but here we are focused on the addition of diffusive transport.

Active matter has become an umbrella term for a range of collective phenomena, but its defining characteristic is incessant local dissipation.
This feature allows for more exotic collective behaviour than is possible in equilibrium systems, such as the separation of self-propelled particles into high- and low-density phases in the absence of any microscopic attractions \cite{catesARCMP2015}.
Phase separation of equilibrating systems (in \eg oil-in-water) is well-captured by phenomenological field theories, so it is natural to ask whether these can be extended to cover phenomena particular to active matter.
Specifically, for a single order parameter field $\phi$ (scalar active matter) it is natural to look for an extension of the \emph{Cahn-Hilliard} equation \cite{cahn1958} (or \emph{Model B} in the nomenclature of \textcite{hohenberg1977}), which has been termed \emph{Active Model B+} (\ambplus) \cite{nardini2017,tjhung2018}.
Its evolution equation reads
\begin{subequations}\label{eq:ambplus}
\begin{equation}
  \partial_t \phi
  =
  \nabla^2{\left(
  \frac{\delta F}{\delta \phi}
  + \frac{\lambda}{2} |\nabla\phi|^2
  \right)}
  - \zeta \nabla \cdot{\left( (\nabla^2 \phi) \nabla\phi \right)}\,,
\end{equation}
where
\begin{equation}
  F[\phi] = \int \dd V \left( f(\phi) + \frac{\kappa}{2} |\nabla\phi|^2 \right)
\end{equation}
\end{subequations}
is the standard free energy functional sufficient to describe an equilibrium system undergoing liquid-gas phase separation \cite{bray2002}.
Here, $f(\phi)$ is the bulk free energy (typically a Ginzburg-Landau form) and $\kappa > 0$ introduces an interface correction.
The coefficients $\lambda$ and $\zeta$ introduce new terms that are \emph{non-integrable} in the sense that they cannot be absorbed into $F$.
This field theory generalises previous field theories for scalar active matter \cite{speck2014,stenhammar2013,wittkowski2014}.
\textcite{nardini2017} suggest that $F$ represents the passive forces, whereas $\lambda$ and $\zeta$ represent the continuous dissipation characteristic of active matter.
This theory exhibits novel patterned microphases \cite{tjhung2018,fausti2021}, critical point properties \cite{caballero2018}, nucleation kinetics \cite{cates2023} and many-body correlations \cite{zheng2023}.

While much of the literature on active matter concerns self-propelled (``motile'') systems, active matter more broadly also encompasses non-motile systems such as reaction-diffusion \eqref{eq:evolution}.
The limiting field theory AmB+ \eqref{eq:ambplus} has been argued purely on symmetry grounds, and so (if correct) should emerge \emph{independent} of such microscopic details.
As such, we might expect to be able to derive AmB+ as a limit of reaction-diffusion \eqref{eq:evolution} with the same symmetries.
Recently, the simpler Model B has been derived for specific models (\ie forms of $\vRflux$) \cite{bergmann2018,bergmann2019}, although a generic and extended derivation (generating $\lambda$ and $\zeta$ terms) is lacking.
Our original motivation to revisit reaction-diffusion has been to explore whether AmB+ could be generically obtained.
If the dissipation terms $\lambda$ and $\zeta$ could be derived, this would potentially enhance the phenomenological understanding of these terms.
Finally, a more direct connection between active matter and reaction-diffusion has been made \cite{kourbane-houssene2018}: the hydrodynamic limit of an active lattice gas reduces to a model that is very similar to reaction-diffusion \eqref{eq:evolution} albeit where $\vRflux$ depends on gradients in $\vrho$ due to particle self-propulsion.

Derivations of \ambplus from microscopic equations of motion have been proposed for active Brownian particles \cite{speck2022,vrugt2023}, active chiral particles \cite{kalz2023} and active switching systems \cite{alston2022}.
In each case the method involves coarse-graining coupled equations of motion in a way that does not isolate the distinct origins of $\lambda$ and $\zeta$.
Their origin may be obscured by the complexity of these models: they involve explicit forces, which is already a significant complication \cite{hansen2013}, and for the self-propelled systems dissipation is intrinsically built into the underlying microscopics rather than being an emergent property of the system coupling to multiple reservoirs.
To simplify these matters, we return to reaction-diffusion \eqref{eq:evolution} as the simplest nontrivial active-matter system.

The goal of the present work is to construct evolution equations for conserved quantities buried in \eqref{eq:evolution} with the perspective to predict and classify patterns even in the presence of many interacting components.
We exploit that the conservation law partitions chemical space into ``slow'' and ``fast'' (reactive) manifolds.
Projection techniques then allow the derivation of evolution equations together with explicit expressions for the coefficients entering these equations.
Non-conserved dynamics in the reactive subspace is crucial but follows the order parameter's dynamics adiabatically, giving rise to corrections at interfaces.
Ultimately we are interested in the \emph{conceptual} bridge between the effective thermodynamics underlying reaction-diffusion, and the top-down field theoretic approach to active systems.

The structure of the paper is as follows.
In \secref{sec:geometry} we introduce the relevant geometry needed to explore chemical space, building on the insights of \Refscite{halatek2018,brauns2020}.
Using this construction, we show in \secref{sec:bulk} that chemical reactions cause an effective bulk free energy to emerge in the translational part of the equation of motion.
The form of this free energy directly relates to the shape of the nullcline (the homogeneous solution of $\vRflux = 0$), with nonlinear terms indicating the system is out of equilibrium.
This emergent free energy dispels any notion that \emph{only} the $\lambda$ and $\zeta$ terms indicate dissipation: even integrable terms can be indicative of chemical driving.
By introducing finite reaction rates in \secref{sec:interface-dynamics}, we show how interface corrections emerge with the integrable $\kappa$ and non-integrable $\lambda$ terms emerging contemporaneously.
We provide a worked example of applying this theory to a minimal model of cell polarisation in \secref{sec:worked-example}.

The sections described above provide an account of how a generalised form of Active Model B emerges from reaction-diffusion, but not the $\zeta$ term in \eqref{eq:ambplus}.
In \secref{sec:patterns} we identify where our framework breaks down as a canididate for where a $\zeta$ term could emerge.
We also describe how the Turing mechanism manifests in reaction-diffusion systems with conservation laws, as well as how our framework realates to more conventional analysis of reaction-diffusion systems.

\section{Geometric preliminaries}
\label{sec:geometry}

\begin{figure}[t]
  \includegraphics[width=\linewidth]{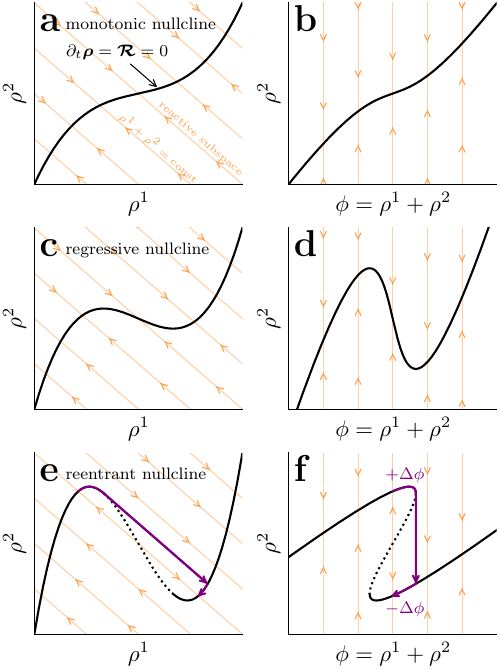}
  \caption{Characteristic phase portraits for two-component systems where the total density $\phi = \rho^1 + \rho^2$ is conserved in the well-mixed limits.
    Chemical fluxes transforms the system homogeneously within the reactive subspace (orange streamlines).
    The homogeneous steady-states occur on the reactive nullcline (black lines) where chemical fluxes vanish; these are categorised into stable (solid black) and unstable solutions (dotted black).
    The right panels (b,d,e) show the same nullclines as the left panels (a,b,c) except plotted with the conserved order parameter $\phi$ along the $x$-axis.
    (a,b) Convex nullclines may have an inflection point, but the tangent vector always points inside the physical quadrant ($\partial_\phi \rho^1(\phi) > 0$ and $\partial_\phi \rho^2(\phi) > 0$ everywhere).
    (c,d) Regressive nullclines have regions where the tangent vector points out of the physical quadrant ($\partial_\phi \rho^1(\phi) < 0$ and/or $\partial_\phi \rho^2(\phi) < 0$).
    These require an inflection point, and the system must have evolved through a plateau where the tangent vector aligns with one of the axes.
    (e,f) Reentrant nullclines exhibit a region of bistability on either side of an unstable branch.
    The system now exhibits a hysteresis loop (purple lines) where the position on the nullcline depends on the history of changes in $\phi$.
    This system is an example of a \emph{cusp catastrophe}.
    }
  \label{fig:phase-portraits}
\end{figure}

\subsection{Central geometric motifs in chemical space}
\label{sec:nullcline}

\begin{figure*}[t]
  \includegraphics[width=\linewidth]{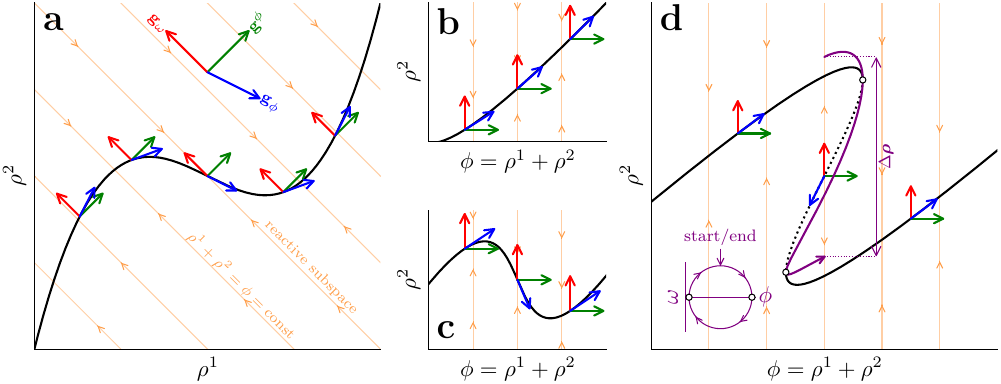}
  \caption{
    Chemical basis vectors for navigating a two-component ($m = 2$) chemical space where density $\phi = \rho^1 + \rho^2$ is conserved.
    $\vec{g}_\phi$ (blue arrows) is the tangent vector the the nullcline (black line), $\vec{g}^\phi = (1, 1)^\top$ (green arrows) the conserved direction and $\vec{g}_\w$ (red arrows) the axis of reactive phase space.
    We show (a) regressive nullcline showing monotonic and regressive regions in the $\rho^1\!\rho^2$-plane, along with zoomed in (b) monotonic and (c) regressive regions following projections onto the $\phi\rho^2$-plane.
    (d) In a reentrant region there are now singular points (white circles) where $\vec{g}_\phi$ and $\vec{g}_\w$ are parallel.
    For (a-c) closed loops in $\phi$ and $\w$ return to their starting $\vrho$.
    For (d) this is not possible in the reentrant region due to hysteresis, as illustrated by a sample path (purple) which is a closed circle in $\phi\w$-space (inset) but generates a jump in $\rho^2$.
  }
  \label{fig:basis-vectors}
\end{figure*}

Our geometric methodology builds on the work of \textcite{brauns2020}.
The central object of study is the \emph{reactive nullcline}.
This nullcline is the set of homogeneous steady-states solving $\partial_t \vrho = \vRflux = 0$.
This solution may result from \eqref{eq:evolution} if the system is continually stirred as in \eg a chemical reactor.
We assume the system exhibits no Hopf bifurcations so that homogeneous steady-states are stationary in time without limit cycles.
The nullcline may have unstable regions, but we assume that the system is confined to the vicinity of the nullcline by the action of chemical fluxes.

In the main text we focus on the case where $\vRflux$ respects a single conservation law, leading to a scalar field theory in a conserved quantity $\phi$.
This generalises to a vector theory in $\vec\phi = (\phi^1, \cdots, \phi^n)^\top$ with multiple conservation laws, discussed in Appendix~\ref{appendix:multiple-conservation-laws}.
For a single law there exists a direction $\vec{g}^\phi$ along which the flux vanishes everywhere $\vec{g}^\phi \vRflux(\vrho) = 0 \; \forall \, \vrho \in \chemicalspace$ such that $\phi \equiv \vec{g}^\phi \vrho$ becomes a conserved quantity.
$\vec{g}^\phi$ is thus a chemical one-form and is a constant everywhere in chemical space.
For convenience, we represent chemical vectors as column vectors and one-forms as row vectors.

Assuming $\vRflux$ is sufficiently smooth, a point solution of $\vRflux = 0$ is extended into a line solution because the conservation law ensures there is at least one direction of marginal stability.
The shape of the resulting nullcline dictates the nature of instabilities in its vicinity, and the resulting field theory.
Regions of the nullcline may be categorised as:
\begin{enumerate}
\item \emph{Monotonic:} the tangent vector to the nullcline can be chosen so that they have only non-negative components.
  That is, they point in the ``physical'' direction in chemical space towards positive concentrations which we will refer to as the \emph{physical orthant}\footnote{An ``orthant'' is the $m$-dimensional generalisation of a quadrant.}.
\item \emph{Regressive:} the tangent vector points outside of the physical orthant, but are everywhere non-orthogonal to the conserved direction.
  The latter condition ensures we can still uniquely resolve the chemical state entirely from the conserved order parameters.
\item \emph{Reentrant:} the regressive regions become so pathological that there is no longer a unique mapping from $\phi$ onto the nullcline.
  This requires the tangent vector to become parallel to the conserved direction at singular points or along a submanifold.
\end{enumerate}
In \Figref{fig:phase-portraits} we illustrate phase portraits for each type for example two-component systems ($m = 2$) with mass conservation $\phi = \rho^1 + \rho^2$ is conserved.

These features of the nullcline are of central importance to the dynamics of a reaction-diffusion system with conservation laws.
\citeauthor{brauns2020} found that linear instability towards pattern formation occurs only when the nullcline becomes regressive for two-component $m = 2$ systems, a result we will generalise to arbitrary $m$ in \secref{sec:bulk}.
Moreover, the nature of this mechanism is modified in the reentrant regime, where the resulting interfaces may become unstable \cite{brauns2020}.

For a two-dimensional chemical space $m = 2$ reentrant nullclines emerge following a \emph{cusp catastrophe} \cite{trong2014}.
In the reentrant region we observe a hysteresis loop because the chemical flux is locally bistable.
Consider the example path (purple line) shown in \Figref{fig:phase-portraits}(e) and (f).
The density of this system is initially decreased by $-\Delta \phi$ in the reentrant zone, pushing the system outside of the stable branch of the nullcline.
The chemical flux then drives the system towards the opposing stable branch.
We return to the original density by increasing the density by $\Delta \phi$, but we remain on the new stable branch at a different chemical concentration.
There has therefore been a net change in $\vrho$ despite the path in $\phi$ being closed.
This hysteresis in the conserved order parameter $\phi$ means it ceases to be a good order parameter so reentrant nullclines are pathological within our current framework.

In the next section we introduce a differential geometric framework that captures the features of regressive nullclines.
Our work thus extends \citeauthor{brauns2020}'s analysis of two-component systems through analytic theory compatible with their more graphical methods.

\subsection{Chemical flux charts the local geometry of chemical space}
\label{sec:chemical-basis}

As the nullcline is the central reference geometry, we must develop a differential geometric framework with which to study it. The purpose of this framework is to resolve components of $\vrho$ into the ``slow'' evolution of $\phi$ and the ``fast'' reactive modes which can be subsequently eliminated.
The intuition underlying our framework is similar to \citeauthor{brauns2020}'s local quasi-steady-state approximation which also took the nullcline as reference.

First, we need a way to characterise the shape of the nullcline via its tangent vectors.
The nullcline is a submanifold of chemical space $\chemicalspace$, which is itself structured by the chemical flux $\vRflux$.
We previously introduced the general reaction-diffusion system \eqref{eq:evolution} with the canonical Cartesian basis $\{\vec e_i\}$ so that $\vrho=\rho^i\vec e_i$.
A Cartesian basis is possible because chemical space is the non-negative orthant of an $m$-dimensional vector space.
Moreover, this basis emerges naturally out of non-interacting stochastic point processes \cite{dean1996,archer2004}, and we assume that the form of the chemical flux is known in this basis $\vRflux = \Rflux^i \vec{e}_i$.

The Cartesian basis is privileged to have a trivial connection $\chemicalgrad$, defined as
\begin{equation}\label{eq:cartesian-connection}
  \LR(\refrho)
  \equiv
  \left. \gradR \right|_\refrho
  =
  \left. \frac{\partial \Rflux^i}{\partial \rho^j} \right|_\refrho \vec{e}_i \otimes \vec{e}^j\,,
\end{equation}
which plays the role of an $m$-dimensional gradient at some point in chemical space\footnote{We reserve $\nabla$ without a subscript for the usual gradient in $d$-dimensional Euclidean space.} $\refrho$.
The one-forms $\{\vec{e}^j\}$ appearing in \eqref{eq:cartesian-connection} satisfy $\vec{e}^j \vec{e}_i = \delta_i^j$.
Note that $\gradR$ is not necessarily symmetric, and so it generally has distinct left and right eigenvectors.
The change in the flux from an infinitesimal change $\delta \vrho$ about point $\refrho$ becomes
\begin{equation*}
  \delta \vRflux
  =
  \LR(\refrho) \delta \vrho
  =
  \frac{\partial \Rflux^i}{\partial \rho^j} \delta \rho^j \vec{e}_i\,.
\end{equation*}
Projecting onto the conserved direction $\vec{g}^\phi$, we find
\begin{equation*}
  \vec{g}^\phi \delta \vRflux
  =
  \vec{g}^\phi \LR(\refrho) \delta \vrho
  =
  0\,,
\end{equation*}
which is zero because $\vec{g}^\phi \vRflux = 0$.
As this must be true for arbitrary $\delta \vrho$, we deduce that the conserved direction $\vec{g}^\phi$ is a left zero eigenvector of $\LR$.
This must be true for any $\refrho$ as $\vec{g}^\phi$ is a global constant of the system.
Note that chemical space is \emph{not} Euclidean: it does not possess translational nor rotational symmetry, and we have no way to quantify \emph{distances} \ie a metric.
Left eigenvectors of $\LR$ are one-forms in our chemical basis.

Suppose there is a special vector $\vec{g}_\phi$ where setting $\delta \vrho = \delta \phi \, \vec{g}_\phi$ for some small $\delta \phi$ leads to no change in flux $\delta \vRflux = 0$, \ie
\begin{equation*}
  \delta \vRflux = \delta \phi \, \left( \LR \vec{g}_\phi \right) = 0\,.
\end{equation*}
This is only satisfied for $\delta \phi \ne 0$ when $\vec{g}_\phi$ is a right zero eigenvector of $\LR$.
This is the tangent vector for the conserved quantity $\phi$, and must coincide with the tangent vector of the reactive nullcline when the flux vanishes at the point $\refrho$.
Excluding singular points where $\vec{g}^\phi$ and $\vec{g}_\phi$ are orthogonal\footnote{These notably occur at turning points of reentrant nullclines which we have already indicated are pathological.}, we set $\vec{g}^\phi \vec{g}_\phi = 1$ so that $\delta \phi = \vec{g}^\phi \delta \vrho$ represents an infinitesimal change in the order parameter.

The most important directions in chemical space are $\vec{g}^\phi$ and $\vec{g}_\phi$ because these capture the curvilinear geometry around the nullcline.
Our construction of the conserved directions in terms of the null space of a linear operator is similar to one used by \citeauthor{falasco2018} in their study of Turing patterns \cite{falasco2018}.
We will assume there is also some convenient basis for the remaining $m-1$ directions (the ``reactive subspace'')\footnote{A fixed basis for the reactive tangent vectors $\{\vec{g}_i\}_{i \ne \phi}$ suffices.
The one-forms are then constructed by the biorthogonality condition $\vec{g}^i \vec{g}_j = \delta_j^i$ for each pair $i,j \in (1, \cdots, m)$.}
This gives a complete set of $m$ one-forms $\{\vec{g}^i\}$ and vectors $\{\vec{g}_i\}$ that describe chemical directions but not distances\footnote{This is normally described as a \emph{conformal geometry}.}.
Directions are sufficient to chart chemical space.
We illustrate the set of directions with example nullclines for a two-component ($m = 2$) mass-conserved system in \Figref{fig:basis-vectors}.

\begin{figure}[t]
  \includegraphics[width=0.8\linewidth]{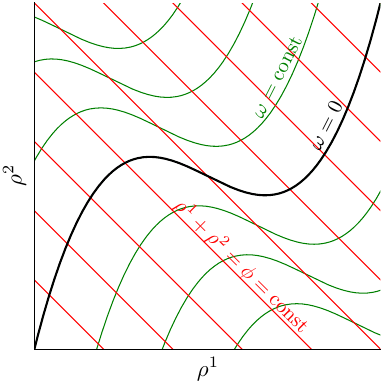}
  \caption{
    Illustration of chemical coordinates for the same two-component system as shown in \Figref{fig:basis-vectors}(a) where density $\phi = \rho^1 + \rho^2$ is conserved and $\w = \rho^2 - \rho^1$ is non-conserved.
    The chemical state is decomposed in our curvilinear basis as $\vrho(\phi, \w) = \vrho_\phi(\phi) + \vwrho(\w)$, where $\vrho_\phi(\phi)$ is obtained by following $\vec{g}_\phi$ (\cf \eqref{eq:scalar-nullcline}) and $\vwrho(\w) = \w \vec{g}_\w$.
    We show the level sets of $\phi$ (red lines) and $\w$ (green).
    The level set with $\w = 0$ is the nullcline (black line), which is simply the $\phi$-axis in our curvilinear basis.
  }
  \label{fig:basis-coords}
\end{figure}

We introduce the coordinates $\vec\x = (\x^1, \cdots, \x^m)^\top = \phi \oplus \vec\w = (\phi, \w^2, \cdots, \w^m)$ where $\vec\w$ are the reactive modes with $\w^1 = 0$ so sums over $\w^i$ exclude $\phi$.
Our coordinates are defined so that directional (covariant) derivatives in the direction of $\vec{g}_i$ in chemical space coincide with partial derivatives with respect to $\x^i$.
This enforces
\begin{subequations}\label{eq:coordinate-scheme}
\begin{equation}\label{eq:differential-basis}
  \vec{g}_i
  =
  \frac{\partial \vrho}{\partial \x^i}
  =
  \vrho_{,i}
\end{equation}
within $U$, using comma notation $(\cdot)_{,i}$ for covariant derivatives in the $\vec{g}_i$ direction.
These derivatives are defined as $\vec{u}_{,i} = \left( \chemicalgrad{\vec{u}} \right) \vec{g}_i$ for $\vec{u} \in \chemicalspace$.
We can invert the definition \eqref{eq:differential-basis} to give $\vrho$ in terms of the coordinates.
This is achieved via contour integration along a path $C$ away from a reference point $\widehat\vrho$:
\begin{equation}\label{eq:rho-contour}
  \vrho
  =
  \widehat\vrho + \int_C \dd\x^i \vec{g}_i\,.
\end{equation}
\end{subequations}
We illustrate the resulting curvilinear basis for an example two-component $m = 2$ system by showing the level sets of conserved $\phi$ and non-conserved $\w$ in \Figref{fig:basis-coords}.
The coordinate chart expressed in \eqref{eq:coordinate-scheme} is valid for a single conservation law because the nullcline is one-dimensional; for multiple conservation laws one would generally have to resort to coordinate expansions within the local neighbourhood of $\widehat\vrho$ (\cf Appendix~\ref{appendix:multiple-conservation-laws}).

\section{Emergent bulk free energy in the limit of infinitely fast reactions}
\label{sec:bulk}

We initially consider the limit where reactions occur infinitely fast.
We do not take this limit explicitly, but rather preclude limit cycles to assume that the system is entirely constrained to the reactive nullcline.
This scenario could correspond to the macroscopic limit where the chemical reactions would dominate diffusion.
For this reason, this limit is indicative of bulk phase behaviour (especially the position of the spinodal) \emph{even in the limit of finite reaction rates}.
The position of the binodal can be affected by coupling to dissipative forces, so we will need more machinery (introduced in \secref{sec:interface-dynamics}) to determine it.

For a single conservation law, we assume that the nullcline is a line that can be parameterised by a scalar $\phi$.
The position along the nullcline is then
\begin{equation}\label{eq:scalar-nullcline}
  \vrho_\phi(\phi)
  =
  \widehat\vrho
  + \int_0^\phi \dd\varphi \, \vec{g}_\phi(\varphi)
\end{equation}
following from \eqref{eq:rho-contour} assuming that the reference state $\widehat\vrho$ exists along the nullcline. Confining the evolution of the system to the nullcline implies $\vrho(\vec{r}, t) = \vrho_\phi(\phi(\vec{r}, t))$.
First-order derivatives are straightforwardly obtained as
\begin{equation*}
  \partial_t \vrho = \left(\partial_t \phi\right) \vec{g}_\phi(\phi)
  \quad \textrm{and} \quad
  \grad{\vrho} = \vec{g}_\phi(\phi) \otimes \grad{\phi}\,.
\end{equation*}
Inserting these into the evolution equation \eqref{eq:evolution} and projecting onto the conserved quantities by contracting with $\vec{g}^\phi$, we obtain the scalar field theory
\begin{equation}\label{eq:scalar-bulk-theory}
  \partial_t \phi
  =
  \Div{\left( \vec{g}^\phi \LD \vec{g}_\phi \nabla\phi \right)}
  \equiv -\Div{\current^\phi}\,.
\end{equation}
This implies a conservative current $\current^\phi = - \grad\mu_\phi$ with effective chemical potential $\mu_\phi(\phi) = \vec{g}^\phi \LD \vrho_\phi$.

In principle, such an effective chemical potential already allows the system to become spatially inhomogeneous with coexisting concentrations along the nullcline given by two values of $\phi$.
The condition for linear instability (the spinodal) is readily obtained from \eqref{eq:scalar-bulk-theory} as
\begin{equation}\label{eq:instability-parameter}
  \epsilon_\finescale \equiv -\vec{g}^\phi \LD \vec{g}_\phi > 0\,.
\end{equation}
This result is compatible with conventional linear stability analysis on the original evolution equation (\cf Appendix~\ref{appendix:linear-stability}).
Several key conditions for instability follow:
\begin{itemize}
\item If $\vec{g}^\phi$ or $\vec{g}_\phi$ are eigenvectors of $\LD$ (with a positive eigenvalue because $\LD$ is positive-definite), then \eqref{eq:instability-parameter} becomes a value proportional to $\vec{g}^\phi \vec{g}_\phi = 1 > 0$ and the homogeneous system is linearly stable.
This must occur if $\LD \propto \Id$ (the identity matrix $\Id$), \ie if all the diffusion coefficients are identical, and so we require asymmetries in diffusion coefficients for phase separation.
Physically this means reactions only become coupled to translations when motion is dependent on chemical composition.
\item A simple way to achieve instability is to tune the values of $\LD$.
By tuning these values, we can achieve negative $\vec{g}^\phi \LD \vec{g}_\phi$ anywhere along the nullcline where $\vec{g}^\phi$ or $\vec{g}_\phi$ are not eigenvectors of $\LD$.
However, for monotonic nullclines (sketched in \Figref{fig:phase-portraits}(a-b)) this requires $\LD$ to have large negative cross-diffusion (``phoretic'') terms (proof in Appendix~\ref{appendix:cross-diffusion}).
As negative cross-diffusion is normally explained as an emergent property of underlying attractions, this scenario is more complex than simple reaction-diffusion of ideal species.
Moreover, it is easy to show that off-diagonal terms in $\LD$ are \emph{non-reciprocal}, meaning they introduce dissipation at the microscopic level.
\item For diagonal $\LD$, instability can be achieved where $\vec{g}_\phi$ points out of the physical orthant, \ie close to an inflection point in the reactive nullcline.
This requires a regressive or reentrant nullcline (sketched in \Figref{fig:phase-portraits}(c-f)).
This generalises \citeauthor{brauns2020}'s instability condition to arbitrary $m$.
\end{itemize}
Here we focus on the last case, where instability occurs without effective attractions and microscopic breaking of time-reversal symmetry.
In principle this case isolates the unadulterated signal of emergent dissipation from bringing together distinct chemical and thermal reservoirs.

Finally, note that we could choose to expand (\cf \eqref{eq:rho-contour}) the chemical potential around $\widehat\vrho$ as
\begin{subequations}\label{eq:scalar-bulk-free-energy-functional}
\begin{equation}\label{eq:scalar-bulk-chemical-potential}
  \mu_\phi
  =
  \vec{g}^\phi \LD \left(
  \frac{\phi}{1!} \hat{\vec{g}}_\phi
  + \frac{\phi^2}{2!} \hat{\vec{g}}_\phi'
  + \frac{\phi^3}{3!} \hat{\vec{g}}_\phi''
  + \cdots
  \right)
  \equiv
  \frac{\delta F_\phi[\phi]}{\delta \phi}\,,
\end{equation}
introducing the effective free energy functional $F_\phi[\phi] = \int_V \dd\vec{r} \, f(\phi(\vec{r}))$ where
\begin{equation}\label{eq:scalar-bulk-free-energy}
  f(\phi)
  \equiv
  \vec{g}^\phi \LD \left(
  \frac{\phi^2}{2!} \hat{\vec{g}}_\phi
  + \frac{\phi^3}{3!} \hat{\vec{g}}_\phi'
  + \frac{\phi^4}{4!} \hat{\vec{g}}_\phi''
  + \cdots
  \right)
\end{equation}
\end{subequations}
is the effective bulk free energy.
The primes indicate normal differentiation \eg $\vec{g}_\phi'(\phi) = \partial_\phi \vec{g}_\phi$.
A hat $\hat{(\cdot)}$ indicates the quantity is evaluated at $\widehat\vrho$, but note that $\vec{g}^\phi$ does not need to carry one because it is a global constant.
The shape of the reactive nullcline entirely dictates the form of this effective bulk free energy.
Thinking of the bulk free energy as the determiner of steady-states for equilibrium systems, it is unsurprising that the nullcline plays a similar role in reaction-diffusion.

We determined above that an inflection point is required for linear instability, so we have to expand the shape of the nullcline at least to cubic order in $\mu$, or quartic in $f$ where it reproduces the familiar $\phi^4$ form characteristic of phase transitions.
This is consistent with the observation that trimolecular reactions are a common motif of reaction-diffusion systems with inhomogeneous steady-states (\eg the Gray-Scott model \cite{gray1983,gray1984}, the Brusselator \cite{prigogine1968} and models for cell polarisation \cite{goryachev2008}).
If the density-dependence of the chemical flux $\vRflux$ is taken solely from the law of mass action, as is normally assumed in reaction-diffusion studies, then trimolecular reactions introduce a cubic concentration-dependence to $\vRflux$ which in turn makes the nullcline cubic.

\section{Limiting field theory for interface formation: finite reaction rates generate an effective surface tension}
\label{sec:interface-dynamics}

\subsection{Evolution equation with deviations from the reactive nullcline}

We define $\vwrho = \w^i \vec{g}_i$ as the non-conserved excess component of $\vrho$, where the tangent vectors $\{\vec{g}_i\}_{i\in\vec\w}$ are some fixed basis in chemical space.
The total concentration is then
\begin{equation*}
  \vrho(\vec{r}, t)
  =
  \vrho_\phi(\vec{r}, t)
  + \vwrho(\vec{r}, t)
  =
  \vrho_\phi(\vec{r}, t)
  + \w^i(\vec{r}, t) \vec{g}_i\,,
\end{equation*}
where $\vrho_\phi$ is the point along the nullcline at the same $\phi$ defined in \eqref{eq:scalar-nullcline}.
The evolution equation \eqref{eq:evolution} becomes
\begin{equation*}
  \left( \partial_t \x^i \right) \vec{g}_i
  =
  \Div{\left(
    \LD \vec{g}_i \otimes \grad{\x^i}
    \right)}
  + \vRflux\,,
\end{equation*}
compactifying notation by using the combined coordinates $\vec\x = \vec\phi \oplus \vec\w$.
The chemical flux vanishes in the $\phi$-direction $\vec{g}^\phi \vRflux = 0$ to conserve $\phi$, leaving
\begin{equation}\label{eq:evolution-finite-reactions}
  \begin{split}
    \partial_t \phi
    &=
    \nabla^2{\left( \mu_\phi + \wmu \right)}
  \end{split}
\end{equation}
where the effective bulk chemical potential $\mu_\phi = f'(\phi)$ is the same as in the limit of infinitely fast reactions \eqref{eq:scalar-bulk-chemical-potential} with the same effective bulk free energy $f(\phi)$ defined in \eqref{eq:scalar-bulk-free-energy}.
The second term introduces a correction to the chemical potential $\wmu = \vec{g}^\phi \LD \nabla^2 \vwrho$ from coupling with non-zero reactive modes.
We will see that this term generates corrections at interfaces, so that bulk phase behaviour is still captured by the previously obtained theory in Sec.~\ref{sec:bulk}.

\subsection{Adiabatic elimination of non-conserved excess concentrations}
\label{sec:simple-adiabatic}

We can project onto changes in non-conserved components by defining the projection operator
\begin{equation*}
  \wproject
  =
  \Id - \vec{g}_\phi \otimes \vec{g}^\phi\,,
\end{equation*}
where $\Id$ is the identity matrix.
Applying this projection operator to the evolution equation \eqref{eq:evolution} gives
\begin{equation}\label{eq:reactive-subspace-evolution}
  \begin{split}
    \partial_t \vwrho
    &=
    \vRflux + \wproject \nabla^2{\left( \vrho_\phi + \vwrho \right)}
  \end{split}
\end{equation}
in this subspace.
We look for a spatio-temporal regime in which the transient evolution of $\vwrho$ vanishes, and $\vwrho$ simply follows the conserved quantity $\phi$ adiabatically where $\vwrho = \vwrho(\phi, \nabla^2 \phi, \cdots)$.

Formally, we would achieve this adiabatic elimination by introducing rescaling of space, time and the chemical concentrations in distinct powers of a small parameter $\epsilon_\finescale$.
For the onset of linear instability, we would take this to be the instability parameter $\epsilon_\finescale = -\vec{g}^\phi \LD \vec{g}_\phi > 0$ from \eqref{eq:instability-parameter}.
By carefully tuning how the various quantities scale with $\epsilon_\finescale$, we can find a regime where an asymptotic series solution for $\vwrho = \vwrho(\phi)$ can be found.
This was the approach of \citeauthor{bergmann2018} \cite{bergmann2018}.
This formal approach gives the result rigorously, but can be opaque to those not already familiar with asymptotic techniques so we proceed more informally.
In Appendix~\ref{appendix:bergmann} we outline the more formal approach at the onset of instability, showing how this gives the same result.

For ease of transparency and to facilitate intuition we proceed here via an informal approach that captures the essential logic of the asymptotic analysis.
In the formal approach we would look for a set of scalings of space, time and the components of $\vrho$ where the dominant behaviour of $\vwrho$ becomes a function of $\phi(\vec{r}, t)$ and its spatial derivatives.
In this regime, the explicit time-evolution of $\vwrho$ appears at higher order and can be neglected.
In the informal approach, we simply set $\partial_t \vwrho = 0$ and solve for $\vwrho$.
This gives
\begin{equation}\label{eq:adiabatic}
  \begin{split}
    \vRflux + \wproject \LD \nabla^2 \vwrho
    &=
    - \wproject \LD \nabla^2 \vrho_\phi\,.
  \end{split}
\end{equation}
At long-times we expect deviations from the nullcline to be small, so we retain only the leading term in the expansion of the flux
\begin{equation*}
  \vRflux = \LR(\phi) \vwrho + \order{|\vwrho|^2},
\end{equation*}
with $\LR(\phi) = \left.\chemicalgrad{\vRflux}\right|_{\vrho=\vrho_\phi(\phi)}$.

Inserting the expansion of $\vRflux$ into \eqref{eq:adiabatic}, and inverting gives solution
\begin{subequations}\label{eq:adiabatic-solution}
\begin{equation}
  \vwrho
  =
  -\left( \Id + \LR^{-1}(\phi) \LD \nabla^2 \right)^{-1}
  \LR^{-1} \LD \nabla^2\vrho_\phi\,.
\end{equation}
The overall concentration $\vrho = \vrho_\phi + \vwrho$ then follows as
\begin{equation}
  \vrho = \left( \Id + \LR^{-1}(\phi) \LD \nabla^2 \right)^{-1}{\vrho_\phi}\,.
\end{equation}
\end{subequations}
As $\LR$ has a null space, we use $\LR^{-1}$ in the restrictive sense to mean the Drazin inverse; this generalised inverse acts solely within the reactive subspace (the row space of $\LR$) so that $\LR^{-1} \LR = \wproject$ and $\LR^{-1} \vec{g}_\phi = \LR \vec{g}_\phi = 0$.
The operator inverse is tricky because of the nonlinearity present in $\LR^{-1} = \LR^{-1}(\phi)$, so it cannot be solved with Green's functions.
Instead, we expand the inverse giving\footnote{This expansion is formally achieved by treating $\LR^{-1} \LD \nabla^2$ as a small operator perturbation, and constructing an asymptotic series for $\vwrho$.
Applying the constraint that $\vec{g}^\phi \vwrho = 0$ excludes any homogeneous solutions solving $\left( \project_\ex + \LR^{-1} \LD \nabla^2 \right)^{-1}(\cdot) = 0$ which would contain a contribution within the conserved subspace.}
the geometric series of operators
\begin{equation}\label{eq:operator-geometric-series}
  \vwrho(\phi)
  =
  \sum_{k=1}^\infty \left(-\LR^{-1}(\phi) \LD \nabla^2\right)^k \vrho_\phi(\phi)\,.
\end{equation}
This expansion of the inverse operator should be valid at large lengthscales, where $\nabla^2$ can be treated as a small perturbation, which is all we need for phase behaviour.
We will continually make use of similar operator expansions, and so we have verified the expansion is consistent with the Green's function solution in the limit where $\LR$ becomes linear in Appendix~\ref{appendix:green}.

The leading excess term in \eqref{eq:operator-geometric-series} is
\begin{equation*}
\begin{split}
  \vwrho
  &=
  - \LR^{-1} \LD \nabla^2 \vrho_\phi
  + \order{\nabla^4}
  \\ &=
  -\LR^{-1} \LD \left(
  \vec{g}_\phi \nabla^2 \phi + \vec{g}_\phi' |\nabla\phi|^2
  \right)
  + \order{\nabla^4}\,.
\end{split}
\end{equation*}
Curvature of the nullcline gives rise to a square-gradient term.
Inserting this term into the conserved equation, we find the leading correction to the effective chemical potential at interfaces is
\begin{subequations}\label{eq:mu-w}
\begin{align}
    \wmu
    &=
    - \left( \vec{g}^\phi \LD \LR^{-1} \right)
    \left( \LD \nabla^2 \vrho_\phi \right)
    + \order{\nabla^4}
    \nonumber \\ &=
    - \left( \vec{g}^\phi \LD \LR^{-1} \right)
    \left( \LD \vec{g}_\phi \nabla^2 \phi + \LD \vec{g}_\phi' |\nabla\phi|^2 \right)
    + \order{\nabla^4}
    \nonumber \\ &=
    - \kappa(\phi) \nabla^2 \phi
    - \frac{\kappa'(\phi) - \lambda(\phi)}{2} |\nabla\phi|^2
    + \order{\nabla^4},
\end{align}
where in the last step we defined
\begin{equation}\label{eq:kappa-lambda}
  \kappa(\phi) =
  \vec\alpha \vec{v}
  \quad \textrm{and} \quad
  \lambda(\phi) = \vec\alpha' \vec{v} - \vec\alpha \vec{v}'
\end{equation}
\end{subequations}
in terms of the point-wise one-form
\begin{subequations}\label{eq:alpha-v}
\begin{equation}
\vec\alpha(\phi) = \vec{g}^\phi \LD \LR^{-1} \wproject
\end{equation}
and the point-wise vector
\begin{equation}
\vec{v}(\phi) = \wproject \LD \vec{g}_\phi
\end{equation}
\end{subequations}
each parameterised by $\phi$.
Note that $\LR^{-1} \wproject = \LR^{-1}$ and $\wproject \wproject = \wproject$, so $\kappa(\phi) = \vec{g}^\phi \LD \LR^{-1} \LD \vec{g}_\phi$ when combined meaning the projection operators are not strictly necessary inside $\vec\alpha$ and $\vec{v}$.
We retain the projections in these quantities to make explicit the fact that the interface terms $\kappa(\phi)$ and $\lambda(\phi)$ arise from a contraction of geometric objects defined in the reactive subspace.

\begin{figure}
  \includegraphics[width=0.9\linewidth]{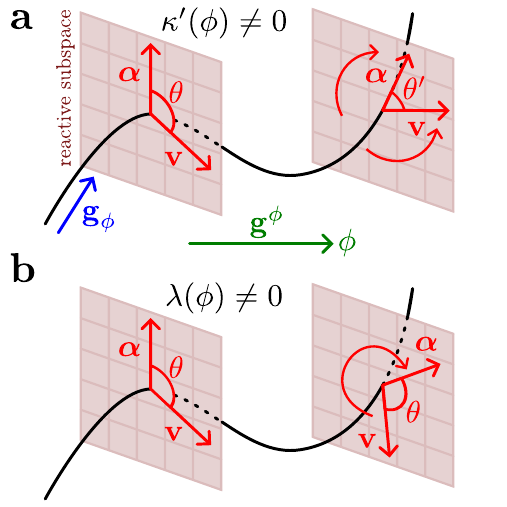}
  \caption{Geometric origin of (a) integrable and (b) non-integrable interface terms in the effective chemical potential $\wmu$ in terms of changes in natural directions $\vec\alpha$ and $\vec{v}$ within chemical space.
    These quantities are all defined in \eqref{eq:mu-w}, and in principle vary along the nullcline (black line) as $\phi$ changes.
  }
  \label{fig:reactive-subspace}
\end{figure}

Note that $\kappa'(\phi) = \vec\alpha' \vec{v} + \vec\alpha \vec{v}'$ is the symmetric counterpart to $\lambda(\phi)$.
$\kappa'(\phi)$ measures the relative changes between $\vec\alpha$ and $\vec{g}_\phi$ (\eg the angle between them), whereas the antisymmetric $\lambda(\phi)$ measures their combined change as we move in the $\phi$-direction (similar to a component of curl).
For $m \ge 3$ this antisymmetric part can be interpreted as a rotation of reactive subspace as we move in the orthogonal $\phi$--direction: this is formally a \emph{torsion}.
In \Figref{fig:reactive-subspace} we illustrate the interpretations of $\kappa'(\phi)$ and $\lambda(\phi)$ as representing how these quantities deform within the reactive subspace (red quantities) as the nullcline (black line) is traversed.

We now wish to understand to what extent the dynamics implied by $\wmu$ can be mapped onto equilibrium physics.

\subsection{Fully integrable limit: Model B}
\label{sec:model-b}

Let us suppose that the evolution equation with finite reaction speeds \eqref{eq:evolution-finite-reactions} can be written entirely in terms of integrable quantities.
That is, we suppose we can write
\begin{equation}\label{eq:model-b}
  \partial_t \phi = \nabla^2{\left( \frac{\delta F}{\delta \phi} \right)}\,,
\end{equation}
with some free energy functional $F = F_\phi + \wF$.
We determined the bulk free energy functional $F_\phi$ previously in \eqref{eq:scalar-bulk-free-energy-functional}, and $\wF$ represents an excess term correcting for interface effects.
In this limit, the binodal can be straightforwardly determined from the Maxwell construction on the effective bulk free energy \eqref{eq:scalar-bulk-free-energy-functional}.

The correction $\wmu(\phi)$ becomes integrable when $\lambda(\phi) = 0$, where the leading interface terms become
\begin{equation*}
  \mu_\ex
  =
  \frac{\delta \wF}{\delta \phi}
  =
  - \kappa(\phi) \nabla^2 \phi - \frac{\kappa'(\phi)}{2} |\nabla\phi|^2\,,
\end{equation*}
with functional
\begin{equation}\label{eq:F-w}
  \wF[\phi] = \int_V \dd\vec{r} \, \frac{\kappa(\phi)}{2} |\nabla\phi|^2\,.
\end{equation}
It is straightforward to show that the $\lambda(\phi) |\nabla \phi|^2/2$ term cannot be written in terms of a free energy functional using arguments originating with \textcite{helmholtz1887}; see also Ref.\ \cite{tonti1969} for an accessible introduction.
Setting $\lambda(\phi) = 0$ implies movements in the function space of $\phi(\vec{r})$ are conservative.
Conversely, $\lambda(\phi) \ne 0$ implies circulation in this function space.

If we were to truncate the evolution equation \eqref{eq:model-b} at $\order{\phi^3}$ in the bulk term $\mu_\phi$ coming from \eqref{eq:scalar-bulk-free-energy-functional} and at leading order in $\wmu$ coming from \eqref{eq:F-w}, then this becomes equivalent to the Cahn-Hilliard equation, or ``Model B'' in the nomenclature of \textcite{hohenberg1977}.
This equation is sufficient to describe phase separation of a conserved quantity, and so provides a suitable starting model for biological condensates.
This would be the dynamical reformulation of the static picture commonly invoked to explain condensates as quasi-equilibrium systems \cite{hyman2014,banani2017}.
Moreover, Model B has been previously derived for specific toy models \cite{bergmann2018,bergmann2019}, extended here to the general case beyond $\order{\phi^3}$.
In this limit the coarse-grained structure of the system is indistinguishable from that of an equilibrating system, so the oil-in-water/Ising paradigm is apt.
The system is not necessarily headed towards microscopic equilibrium, as non-equilibrium driving forces may be built into $\vRflux$ underlying the coefficients $f(\phi)$ and $\kappa(\phi)$; however, such dissipative forces cannot be inferred from evolution of the order parameter alone and the system minimises an effective free energy.

\subsection{Non-integrable generalisation: Active Model B}
\label{sec:active-model-b}

Now we consider the more general case where $\lambda(\phi) \ne 0$; this term has not been previously derived for a conserved reaction-diffusion model to the best of our knowledge.
We decompose the interface contribution to the chemical potential as
\begin{equation*}
  \wmu
  =
  \frac{\delta \wF}{\delta \phi}
  + \frac{\lambda(\phi)}{2} |\nabla\phi|^2
  + \order{\nabla^4}\,,
\end{equation*}
where $\lambda(\phi)$ is given in \eqref{eq:kappa-lambda}.
The evolution equation is now
\begin{equation}\label{eq:active-model-b}
  \partial_t \phi
  =
  \nabla^2{\left( \frac{\delta F}{\delta \phi} + \frac{\lambda(\phi)}{2} |\nabla\phi|^2 \right)}
  + \order{\nabla^6}\,,
\end{equation}
where the free energy $F = F_\phi + \wF$ is the same as before in \eqref{eq:model-b}.
Truncating the $\mu_\phi(\phi)$ term in the evolution equation at order $\phi^3$, and the interface terms $\kappa(\phi)$ and $\lambda(\phi)$ at leading order reduces \eqref{eq:active-model-b} to so-called ``Active Model B'' \cite{stenhammar2013,wittkowski2014}.
This model was introduced to describe the macroscopic behaviour of motile active systems, which generically feature dissipation (even in the steady-state) due to inherent driving.
In our case, the chemical flux $\vRflux$ will generally describe non-equilibrium chemical dynamics unless it is finely tuned, so it is unsurprising that we can make a correspondence with active matter.

The position of the binodal in this case is more subtle than for the $\lambda(\phi) = 0$ case.
It can be determined using the techniques of \Refscite{solon2018,tjhung2018}, by considering interfacial profiles with a planar symmetry and constructing an integrating factor.

\begin{figure}
  \includegraphics[width=0.9\linewidth]{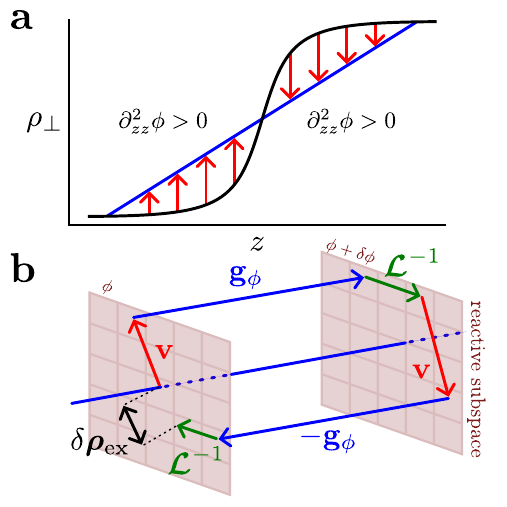}
  \caption{
    (a) Kink in $\phi \to \delta \phi$ along the $z$ direction results in deviations of some perpendicular component $\rho_\perp$ (black line) away from the nullcline (blue line) which is a straight line for small enough $\delta \phi$.
    The chemical flux vectors (red arrows) attempt to restore the perpendicular component back to the nullcline (but these may be balanced by diffusive currents).
    (b) Sketch of non-reciprocal reaction pathway in chemical space.
    The straight nullcline along which the kink in $\delta \phi$ occurs is shown penetrating adjacent reactive subspaces.
    Non-reciprocal reaction pathway is overlaid (thick arrowed lines).
    Application of the first $\LR^{-1}$ determines the change in excess density $\vrho_\ex$ at $\phi + \delta \phi$ from action of the chemical flux.
    The second $\delta \vRflux$ is the chemical flux in response to this excess density.
  }
  \label{fig:lambda-kink}
\end{figure}

Let us close by providing a more physical interpretation of $\lambda(\phi)$ in the specific context of reaction-diffusion.
For small changes in $\phi$ the nullcline will be effectively a straight line, so we can ignore curvature within small regions of a phase boundary.
For illustration purposes we consider an infinitesimal kink in $\vrho \to \vrho + \delta\vrho$ from a small change $\phi \to \phi + \delta \phi$ across a phase boundary (\eg around onset of phase separation) as sketched in \Figref{fig:lambda-kink}(a).
From \eqref{eq:kappa-lambda}, we rewrite $\lambda(\phi) = -\vec{g}^\phi \LD \delta\vrho_\ex$ where
\begin{align}
\label{eq:kink-excess}
  \delta \vwrho
  &\equiv \LR^{-1} \vec{v}' - (\LR^{-1})' \vec{v}
  \\ \nonumber &=
  \lim_{\delta \phi \to 0}
  \frac{1}{\delta \phi} \left(
  \LR^{-1}(\phi) \vec{v}(\phi + \delta \phi) - \LR^{-1}(\phi + \delta \phi) \vec{v}(\phi)
  \right)
\end{align}
is an excess concentration per magnitude of the kink $\delta \phi$.
The vector $\vec{v} = \LR \left( \LR^{-1} \LD \vec{g}_\phi \right)$ is the chemical flux in response to the kink in $\phi$ because $\vwrho = \LR^{-1} \LD \vec{g}_\phi$ per unit $\nabla^2 \phi$ (we can ignore $|\nabla\phi|^2 \ll 1$ terms when $\delta \phi$ is small).
Subsequent action with $\LR^{-1}$ maps this infinitesimal flux back onto an excess concentration $\vwrho$.
These two operations happen twice in \eqref{eq:kink-excess}, alternately interrupted by movement along $\vec{g}_\phi$ due to a gradient in $\phi$; the closed circuit in \eqref{eq:kink-excess} then represents a novel reaction pathway made possible by an interface where diffusion explicitly couples to the reactions.
Non-vanishing $\delta \vwrho \ne 0$ indicates a net change along this emergent reaction pathway.
We sketch the circuit in \Figref{fig:lambda-kink}(b).
Contracting $\delta\vrho_\ex$ with $\vec{g}^\phi \LD$ to give $\lambda(\phi)$ (up to a minus sign) simply determines the effect this excess contribution has on transport of $\phi$.
We can rewrite $\lambda(\phi)$ as the operator\footnote{We need to write $\vec{g}_\phi' = -\LR^{-1} \LR' \vec{g}_\phi$ from differentiating $\LR \vec{g}_\phi = 0$, and similarly $\LR (\LR^{-1})' = \project_\ex^{\;\prime} - \LR' \LR^{-1}$ from differentiating $\LR \LR^{-1} = \project_\ex$.}
\begin{equation*}
\lambda(\phi)
=
\vec{g}^\phi \LD \LR^{-1} \left( \project_\ex^{\;\prime} \LD + \LD \LR^{-1} \LR' - \LR' \LR^{-1} \LD \right) \vec{g}_\phi\,.
\end{equation*}
The operator $\project_\ex^{\;\prime} \LD + \LD \LR^{-1} \LR' - \LR' \LR^{-1} \LD$ provides an alternative representation of the emergent reaction pathway.

\begin{figure}[t]
  \includegraphics[width=\linewidth]{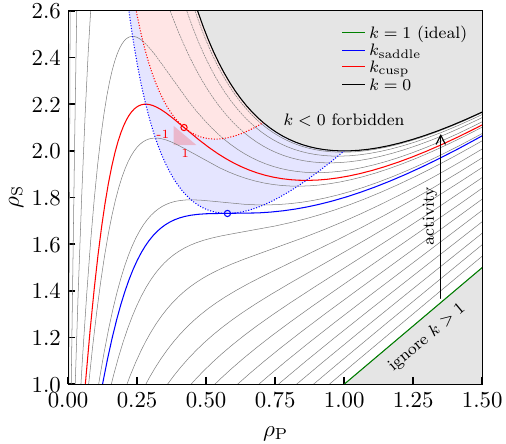}
  \caption{Nullclines (lines) and phase behaviour (shaded regions) for a two-component toy model of cell polarisation (given in \eqref{eq:cell-polarisation-flux}) with a single model parameter $k \in [0, 1]$.
    In the blue region where $k < k_\mathrm{saddle}$ the system \emph{may} be linearly unstable to phase separation, though the exact position of the spinodal lines depends on the diffusion coefficients; the borders of this region (blue dotted) form the spinodal lines in the limit $D_\product / D_\substrate \to 0$.
    For $k < k_\mathrm{cusp}$ the dynamics around the nullcline becomes locally bistable in the reentrant region (red shaded area); outside of this region and/or for $k > k_\mathrm{cusp}$ the chemical flux is monostable.
  }
  \label{fig:example-phase-diagram}
\end{figure}

\section{Worked example: cell polarisation}
\label{sec:worked-example}

\begin{figure}[t]
  \includegraphics[width=\linewidth]{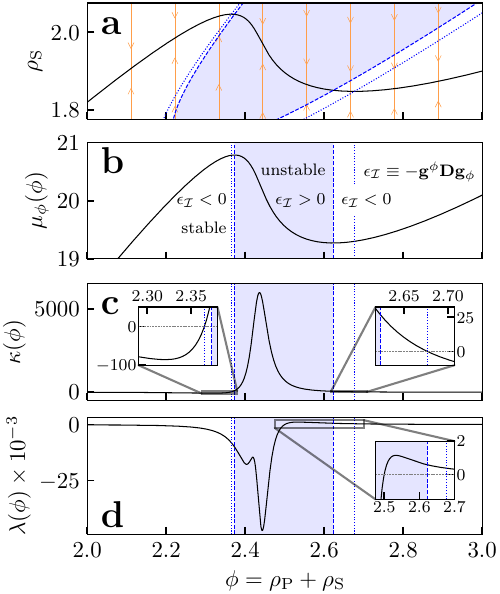}
  \caption{Parameters for the limiting scalar field theory for the two-component cell polarisation model introduced in \Figref{fig:example-phase-diagram} (and given in \eqref{eq:cell-polarisation-flux}) with parameters $k = 0.07$, $D_\product = 1$ and $D_\substrate = 10$.
    (a) Phase diagram showing regions of stable homogeneous solutions (white fill), linearly unstable region (blue fill) and the spinodal (dashed line).
    We also show the limiting spinodal (dotted line) where the coexisting region reaches a maximum area in the limit $D_\product / D_\substrate \to 0$.
    The chemical flux pushes the system towards the nullcline (orange lines).
    (b) Effective chemical potential along the nullcline has a cubic shape justifying a $\phi^4$ model.
    Also shown are the coefficients of (c) integrable $\kappa(\phi)$ and (d) non-integrable $\lambda(\phi)$ interface corrections to the effective chemical potential along the nullcline.
    Note that $\kappa(\phi)$ is positive in the phase-coexisting domain, and negative where only homogeneous phases are possible (insets).
    By constrast, $\lambda(\phi)$ can have either sign in each domain.
  }
  \label{fig:example-parameters-normal}
\end{figure}

At this stage it is instructive to work through a specific example to illustrate how one performs calculations within our differential geometric framework.
We omit the final coefficients if they involve long algebraic expressions.

We adapt a two-component ($m=2$) model for cell polarisation from \textcite{mori2008}.
This models the nonlinear catalysis of a substrate $\substrate$ into a product $\product$ monomer.
We write $\vrho = (\rho_\product, \rho_\substrate)^\top$, and the process is \ce{$\substrate$ <=> $\product$} with chemical flux
\begin{equation}\label{eq:cell-polarisation-flux}
  \vRflux
  =
  \left[
  \left( k + (1 - k) \frac{\rho_\product^2}{1 + \rho_\product^2} \right) \rho_\substrate
  - \rho_\product
  \right]
  \begin{pmatrix}1 \\-1\end{pmatrix}
\end{equation}
where $k \in [0, 1]$.
This is the same as the chemical flux used in \Refcite{mori2008}, except we have a $(1-k)$ prefactor for the second term\footnote{This model emerges from coarse-graining a five-component model which additionally contains: a fuel monomer $\fuel$, depleted fuel $\spentfuel$ and an intermediate complex $complex$.
These react in the following main pathway:
\[\ce{$\substrate$ + 2$\product$ + $\fuel$ <=>[$\tfrac{1}{2}$][1] $\complex$ ->[1] 3$\product$ + $\spentfuel$}\,.\]
The difference in chemical potential between $\fuel$ and $\spentfuel$ makes the second step irreversible.
This reaction pathway is sustained by additional chemostatting steps \ce{$\fuel$ <=>[$\infty$][$\infty$] $\spentfuel$} (so fuel concentrations are constant) and \ce{$\substrate$ <=>[k][1] $\product$}.
Eliminating the concentrations of $\fuel$, $\spentfuel$ and $\complex$ (via an adiabatic approximation) leads to the stated model.}.
In addition to simplifying resulting analytical expressions, the $(1 - k)$ prefactor allows us to recover an ideal limit when $k = 1$ and the nullcline becomes completely linear\footnote{In the previous footnote we introduced $k$ as a rate of chemostating, so $k = 1$ implies there is no chemical potential difference between $\substrate$ and $\product$.
This would \emph{seem} to be an equilibrating system from the top-down perspective of the field theory.
However, the underlying system is still driven as it contains an irreversible step converting $\fuel \to \spentfuel$.}.
This limit is not necessarily microscopically in equilibrium, but it is indistinguishable from a passive ideal gas in its macroscopic evolution equation.
Decreasing the parameter $k$ from unity (loosely) increases the chemical affinity biasing the catalysis of \ce{$\substrate$ -> $\product$}.
The conservation law for \eqref{eq:cell-polarisation-flux} is $\phi = \rho_\substrate + \rho_\product$ in the direction of $\vec{g}^\phi = (1, 1)^\top$.

In \Figref{fig:example-phase-diagram} we sketch the nullclines for \eqref{eq:cell-polarisation-flux} vs the $k$ parameter overlayed by the phase diagram.
The nullclines are found by solving $\vRflux(\vrho_\phi) = 0$, giving
\begin{equation*}
  \hat\rho_\substrate = \frac{1 + \hat\rho_\product^2}{k + \hat\rho_\product^2} \hat\rho_\product
\end{equation*}
where hats $\hat{(\cdot)}$ indicate the quantity is evaluated on the nullcline so $\vrho_\phi = (\hat\rho_\substrate, \hat\rho_\product)^\top$.
Varying the parameter $k$, there are three regions corresponding to the distinct nullcline geometries sketched in \Figref{fig:phase-portraits}.
Defining $k_\mathrm{saddle} = 1/9$ (nullcline sketched with blue line in \Figref{fig:example-phase-diagram}) and $k_\mathrm{cusp} = 1/17$ (red line), these regions are:
\begin{enumerate}
\item $k > k_\mathrm{saddle}$: the nullcline is monotonic and the system is homogeneously stable.
\item $k_\mathrm{cusp} < k < k_\mathrm{saddle}$: the nullcline is regressive, possessing a region where $\tfrac{\dd \hat\rho_\substrate}{\dd \hat\rho_\product} \le 0$ (blue shaded region in \Figref{fig:example-phase-diagram}).
The chemical dynamics around the nullcline in this region remain monostable, but the system can phase separate when the inequality \eqref{eq:instability-parameter} is satisfied.
The blue region shaded in \Figref{fig:example-phase-diagram} represents the \emph{maximum} boundaries of the spinodal (achieved in the limit $D_\product / D_\substrate \to 0$), whereas the realised spinodal depends on the coefficients of $\LD$.
The nullcline necessarily has an inflection point.
\item $k < k_\mathrm{cusp}$: the nullcline develops a reentrant region where $\tfrac{\dd \hat\rho_\substrate}{\dd \hat\rho_\product} \le -1$ (red shaded region in \Figref{fig:example-phase-diagram}).
The chemical dynamics around the nullcline are locally bistable inside this region, and monostable everywhere else.
The nullcline ceases to be a good reference geometry in the bistable region making it pathological within our framework; outside of the bistable region our framework still applies.
\end{enumerate}

To evaluate the coefficients appearing in our active field theory we need the tangent vectors and the linear expansion of the chemical flux.
The tangent vector is determined by taking the gradient of $\vrho_\phi$ in chemical space, giving
\begin{equation*}
  \vec{g}_\phi
  =
  \frac{1}{G} \left(
  1,
  \frac{k + \left( 3 k - 1 \right) \hat\rho_\product^2 + \hat\rho_\product^4}{(k + \hat\rho_\product^2)^2}
  \right)^\top
\end{equation*}
with normalisation constant $G$ ensuring $\vec{g}^\phi \vec{g}_\phi = 1$.
The reactive subspace then occurs along the line $\vec{g}_\w \propto (1, -1)^\top$ which are the level sets of $\phi$, with $\vwrho = \w \vec{g}_\w = 0$ on the nullcline.
The linear chemical flux about the nullcline is $\LR \vwrho$ where
\begin{equation*}
  \LR
  =
  \eval \wproject
  =
  \eval
  \left( \Id - \vec{g}_\phi \otimes \vec{g}^\phi \right)\,,
\end{equation*}
with the eigenvalue $\eval$ obtained from diagonalising $\chemicalgrad \vRflux$ at points along the nullcline:
\begin{equation*}
  \eval
  =
  -2 + \frac{3 - k}{1 + \hat\rho_\product^2}
  - \frac{2 k}{k + \hat\rho_\product^2}\,.
\end{equation*}
Its inverse within the reactive subspace is simply
\begin{equation*}
  \LR^{-1}
  =
  \frac{\wproject}{\eval}
  =
  \frac{1}{\eval}
  \left( \Id - \vec{g}_\phi \otimes \vec{g}^\phi \right)\,.
\end{equation*}
We will initially concentrate on the case $k > k_\mathrm{cusp}$ where the nullcline is homogeneously stable, \ie $\eval < 0$ everywhere.
We will return to consider the case $k < k_\mathrm{cusp}$ at the end of this section.

\begin{figure}[t]
  \includegraphics[width=\linewidth]{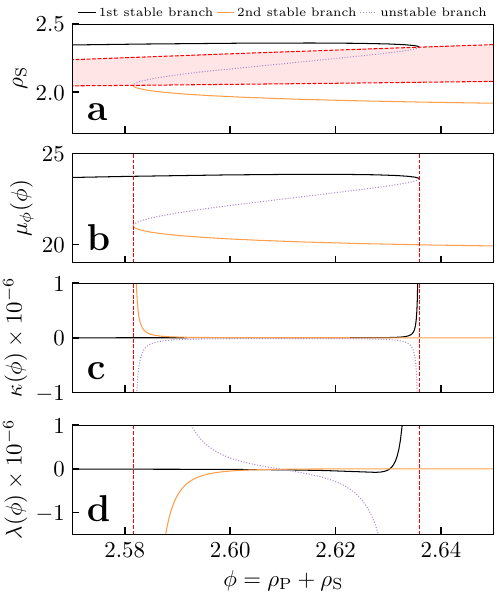}
  \caption{
    Parameters for our limiting scalar field theory with $D_\product = 1$, $D_\substrate = 10$ and $k = 0.05$ for the cell polarisation model specified by \eqref{eq:cell-polarisation-flux}; this value of $k$ is within the regime where the nullcline becomes reentrant.
    This parameterisation is valid along the stable branches (solid lines) up to the turning points, and invalid along the unstable branch but can be analytically continued there (dotted purple lines)
    (a) Chemical dynamics around a reentrant nullcline (lines) is locally bistable (red region).
    (b) Effective chemical potential along the nullcline can be analytically continued in straightforwardly.
    The coefficients of (c) integrable $\kappa(\phi)$ and (d) non-integrable $\lambda(\phi)$ interface corrections to the effective chemical potential along the stable branches of the nullcline asymptote to $+\infty$ and $\pm\infty$ respectively at the turning points.
    Curiously, analytic continuation of $\kappa(\phi)$ into the unstable branch is \emph{negative}, and asymptotes to $-\infty$ at the turning points.
  }
  \label{fig:example-parameters-reentrant}
\end{figure}

We illustrate parameters of the field theory in regions where the nullcline is regressive but not reentrant in \Figref{fig:example-parameters-normal}.
The bulk part of the effective chemical potential \eqref{eq:scalar-bulk-chemical-potential} is
\begin{equation*}
  \begin{split}
  \mu_\phi
  =
  \vec{g}^\phi \LD \vrho_\phi
  &=
  D_\substrate \frac{1 + \hat\rho_\product^2}{k + \hat\rho_\product^2} \hat\rho_\product
  + D_\product \hat\rho_\product\,.
  \end{split}
\end{equation*}
Changing variables from $\hat\rho_\product$ to density $\phi$ and expanding to cubic order in $(\phi - \phi_0)$ where $\phi_0 = \phi(t=0)$ is the initial condition gives the familiar Ginzburg-Landau chemical potential \eqref{eq:scalar-bulk-chemical-potential}.
We sketch the chemical potential in \Figref{fig:example-parameters-normal}(b) which has the characteristic cubic form of $\phi^4$ theory; this chemical potential is a simple transformation of the cubic shape of the nullcline in \Figref{fig:example-parameters-normal}(a).
The integrable (``passive'') interface correction is
\begin{equation*}
  \begin{split}
    \kappa(\phi)
    &=
    \vec{g}^\phi \LD \LR^{-1} \LD \vec{g}_\phi,
    \\ &=
    \frac{
      k + (3k - 1) \hat\rho_\product^2 + \hat\rho_\product^4
    }{
      \left(1 + \hat\rho_\product^2 \right)^2
    }
    \frac{(D_\substrate - D_\product)^2}{\eval^3}\,.
  \end{split}
\end{equation*}
For comparison with Active Model B (\eqref{eq:ambplus} with $\zeta = 0$), we would only retain the leading (constant) contribution to $\kappa(\phi)$ at an initial reference state so $\kappa'(\phi) = 0$.
The non-integrable term $\lambda(\phi)$ is obtainable in the same way, but we omit its more complicated expression.
We sketch these interface terms in \Figref{fig:example-parameters-normal}(c-d).
We find that $\kappa(\phi) > 0$ inside the region where the system is linearly unstable towards phase separation (which is consistent with Model B dynamics), and $\kappa(\phi) < 0$ in the stable regions.
There do not appear to be any constraints on the sign of $\lambda(\phi)$.
As $k \to 1$ in the ideal case, we find
$\lambda \to 0$ as expected.
This case cannot be distinguished from an equilibrating ideal gas from the top-down reaction-diffusion equation \eqref{eq:evolution} alone; a bottom-up approach would be needed to determine whether there is dissipation.

Finally, let us consider briefly the case $k < 1/17$ where the nullcline becomes reentrant.
In this case there are two stable branches terminating in singular turning points which are joined by an unstable branch.
Our theory remains valid along the stable branches far from the turning points, where the dynamics is uninteresting (\ie stable).
We sketch this geometry for a particular value of $k$ in \Figref{fig:example-parameters-reentrant}(a).
The dynamics approaching the turning points (or between them) is pathological within our theory, as we have only retained linear terms in the chemical flux.
We can nonetheless analytically continue our parameters to see how they behave approaching the pathological regime.
We sketch the bulk and interface terms in the effective chemical potential in \Figref{fig:example-parameters-reentrant}(b-d).
Curiously, we find that the interface terms asymptote to $\pm \infty$ as the turning points are approached from the stable branches; this is simply a consequence of the eigenvalue of $\LR$ vanishing at the turning point, and so its inverse will diverge at these points.
The interface terms change signs when analytically continued into the unstable branch, asymptoting to $\mp \infty$ at the turning points.
This singular behaviour illustrates the breakdown of our theory for reentrant nullclines, indicating a breakdown of our asymptotic series solution.

\section{Connections with conventional approaches and pattern formation}
\label{sec:patterns}

Linear stability analysis is a common starting point for detecting pattern-forming instabilities.
By considering the evolution of perturbations $\delta\vrho \equiv \vrho - \widehat\vrho$ about some reference homogeneous state $\widehat\vrho$:
\begin{equation*}
\partial_t \widetilde{\delta\vrho} = \left( \widehat\LR - q^2 \LD \right) \widetilde{\delta\vrho}\,,
\end{equation*}
where a tilde $\widetilde{(\cdot)}$ represents the Fourier transform.
The operator $\mathbf{L} = \widehat\LR - q^2 \LD$ develops a zero eigenvalue at the Turing bifurcation.
In principle instabilities could occur purely along non-conserved components.
As we are only interested in the novelty of conservation laws on reaction-diffusion, we must project onto $\phi$.
A minimal model for a Turing instability along the conserved mode is
\begin{equation*}
\partial_t \widetilde\phi = -q^2 \left( a - \kappa q^2 + \eta q^4 \right) \widetilde\phi
\end{equation*}
We require $a > 0$ and $\eta > 0$ to ensure stability at small and large $q$.
The first requirement implies that the nullcline is monotonic.
The homogeneous state develops a Turing instability at some finite $q > 0$ for $\kappa < -\sqrt{4 a \eta}$.
We can now generalise this criterion as our general solution contains within it all terms of linear stability analysis.
In particular, we have the first few terms
\begin{align*}
  a &= \vec{g}^\phi \LD \widehat{\vec{g}}_\phi \, \\
  \kappa &= \vec{g}^\phi \LD \widehat{\LR}^{-1} \LD \widehat{\vec{g}}_\phi \,, \\
  \eta &= \vec{g}^\phi \LD \widehat{\LR}^{-1} \LD \widehat{\LR}^{-1} \LD \widehat{\vec{g}}_\phi \,.
\end{align*}
These are simply the leading terms in the expansion of our adiabatic solution \eqref{eq:adiabatic-solution} with constant $\LR$.
Conserved modes therefore become marginally stable when
\begin{equation*}
  \vec{g}^\phi \LD \left( \Id - q^2 \widehat{\LR}^{-1} \LD \right)^{-1} \widehat{\vec{g}}_\phi = 0\,.
\end{equation*}
This can be inverted to design systems with Turing bifurcations at specific wavelengths.
The coefficients $\{a, \kappa, \eta\}$ only become linearly independent if $m \ge 3$ as a consequence of the Cayley-Hamilton theorem, so at least $3$ species are required for this class of patterns.

Our analysis is connected with weakly nonlinear approaches.
Going beyond linear stability analysis requires exploiting the zero eigenvector of $\mathbf{L}$ at the Turing bifurcation.
The nonlinear evolution of amplitudes along the marginally stable mode can then be constructed by Fredholm alternative \cite{cross1993,desai2009}.
The Ginzburg-Landau equation is an example of an amplitude equation that can emerge \cite{kuramoto1975,kuramoto1976,cross1993}.
Our approach exploits the fact that just $\widehat\LR$ has a nullspace due to the conservation law.
This property allows construction of a nonlinear solution without as austere a restriction to the onset of instability.
\citeauthor{bergmann2018} recognised this in constructing a solution near the spinodal, and we have here exploited it beyond onset.
In Appendix~\ref{appendix:bergmann} we detail the weakly nonlinear analysis of \citeauthor{bergmann2018} and show its equivalence to our approach of previous sections.

Our approach will not capture patterns where the expansion of our adiabatic solution \eqref{eq:adiabatic-solution} as a geometric series \eqref{eq:operator-geometric-series} does not converge.
In this case the leading (linear) solution can still be represented \emph{non-locally} in terms of the Green's function of the screened poisson equation, but this cannot be reduced to a local representation.
The effective free energy becomes intrinsically non-local.
Nonlocal terms of this kind have been seen previously in pattern-forming systems \cite{ohta1989,petrich1994,drazer1997}.
We have shown that the Green's function solution corresponds to the linear case (Appendix \ref{appendix:green}), but this can be formally broken by introducing a singular perturbation.
This occurs where the operator $\LR^{-1} \LD$ splits into an $\order{1}$ component and a large $\order{\epsilon^{-1}}$ component which violates the assumed sizes of sequential terms in the geometric series.
We note several prominent pattern-forming models with conservation laws feature a small parameter, either in the diffusion coefficients \cite{holmes2012,halatek2018,gai2020} or in the flux $\vRflux$ \cite{murray2017}.

\section{Perspective}
\label{sec:perspective}

We have derived a limiting scalar field theory \eqref{eq:active-model-b} describing the evolution of a reaction-diffusion system respecting a conservation law (with a straightforward generalisation to multiple conserved quantities in Appendix~\ref{appendix:multiple-conservation-laws}).
This field theory corresponds to a generalisation of (Active) Model B \cite{stenhammar2013,wittkowski2014} (\eqref{eq:ambplus} with $\zeta = 0$) that does not truncate at $\order{\phi^3}$.
The emergent bulk free energy / chemical potential is obtained from the shape of the nullcline (or a more general contour from Appendix~\ref{appendix:arbitrary-reference}), and without truncation it is in principle arbitrarily accurate approaching the binodal; this analytic result is broadly compatible with previous geometric constructions \cite{brauns2020}.
Without truncation we also obtained point-wise concentration-dependent coefficients of the interface terms \ie $\lambda \to \lambda(\phi)$ and $\kappa \to \kappa(\phi)$.

Previous attempts to derive a limiting scalar field theory for reaction-diffusion have not resulted in a $\lambda(\phi)$ term.
Most notably, the arguments of \textcite{bergmann2018} assumed particular asymptotic scalings focusing on onset dynamics which push any non-potential term to higher-order in field gradients; by contrast, our work focuses on longer-time dynamics.
Our theory remains valid for as long as the homogeneous solution (along the nullcline) remains a good reference state.
Our framework can be straightforwardly generalised to more general reference contours (\ie we do not strictly have to use the nullcline) allowing it to become asymptotically exact approaching the steady-state.
We have relocated this generalisation to Appendix~\ref{appendix:arbitrary-reference} to keep our main argument focused on the nullcline, but we note that the resulting field theory \eqref{eq:active-model-b} remains valid beyond onset (albeit with more complicated coefficients) and approaching the steady-state.

Our theory results in a whole hierarchy of non-integrable terms in the emergent chemical potential, in which the square-gradient $\lambda$ term conjectured in Active Model B \cite{stenhammar2013,wittkowski2014} is the leading contribution.
Even when it cannot be mapped onto an equilibrating system, it retains a resemblance to equilibrium systems in that the currents are represented by local functions and vanish in the steady-state.
In this sense $\lambda(\phi)$ introduces non-conservative (yet still local) effective forces, emerging from path-dependence of the free energy through function space.
The meaning of this path-dependent term can be interpreted in multiple equivalent ways:
\begin{itemize}
\item \emph{Thermodynamically} indicating violation of a variational principle \ie explicit dissipation.
\item \emph{Functionally} as circulation in $\phi$'s function space.
\item \emph{Geometrically} emerging from torsion in the reactive subspace: specifically from rotations of the natural directions $\vec{\alpha}$ and $\vec{v}$ defined in \eqref{eq:kappa-lambda}.
\item \emph{Chemically} as a novel reaction pathway emerging at phase boundaries.
\end{itemize}
As established in \secref{sec:bulk}, the nonlinear terms in the bulk free energy are themselves dissipative.
It is clear that dissipation can manifest in multiple ways, only some of which are visibly path-dependent in the spatio-temporal regime where we can project the system's high-dimensional dynamics onto a single order parameter $\phi$.
In other words, the true thermodynamics is obscured by coarse-graining.
As such, integrable terms in the effective free energy are ambiguous: from a top-down perspective, we cannot say whether they are passive or not.
The $\lambda(\phi)$ term captures the leading path-dependent behaviour, and is \emph{necessarily} dissipative.

The main takeaway for cell biologists is that we can achieve all the same phenomenology of oil-in-water phase separation (\ie the dominant paradigm in condensates) \emph{without} effective attractions \cite{hyman2014,banani2017,weber2019,falahati2019}.
A recent example has shown phase separation in enzymes in the absence of equilibrium interactions \cite{cotton2022}.
That is not to say that forces are unimportant in biology (they clearly are), but that they may tell an incomplete story.

We had hoped that our nonlinear treatment would provide insight into the mysterious $\zeta$ term, but our approach appears too limited.
The $\zeta$ term appearing in \ambplus \eqref{eq:ambplus} introduces an inherently non-local component to the chemical potential \cite{tjhung2018}.
We have noted that non-local terms are expected precisely where our theory would fail to converge \cite{ohta1989,petrich1994,drazer1997}; study of these system could lead to understanding of how the $\zeta$ emerges and its thermodynamic meaning.
Finally we remark that going beyond ideal diffusion with particle-based forces introduces non-local terms in the free energy functional \cite{lu1985}.
Stable stationary patterns have been seen in chemically reactive generalisations of Model B \cite{glotzer1995} and dynamical density functional theory \cite{alston2022}.
The latter case even reduced to \ambplus.

\begin{acknowledgments}
  JFR gratefully acknowledges financial support from the Alexander von Humboldt foundation.
  JFR wishes to thank Francesco Turci for several informative conversations concerning this work and its connections with active matter, and Bob Evans for pointing out that equilibrium liquid-gas interfaces require a non-local free energy functional.
  JFR would also like to thank Tobias Galla for his inspiring lecture, boldly titled `How the Zebra Gets Its Stripes'; a decade of joyful reminiscence ultimately culminated in this investigation.
  TS acknowledges financial support from Deutsche Forschungsgemeinschaft through SFB 1551 (grant no.\ 464588647).
\end{acknowledgments}

\appendix

\section{Multiple conservation laws}
\label{appendix:multiple-conservation-laws}

\subsection{Differential geometry}

In the main text we focused on the case of a single conservation law for simplicity.
Our framework naturally extends to multiple conservation laws.
For $n < m$ conservation laws, the nullcline becomes an $n$-dimensional submanifold and we develop a set of $n$ conserved quantities $\vec\phi = (\phi^1, \cdots, \phi^n)^\top$ in some basis.
For $n > 1$ there is no unique choice of $\vec\phi$.
The null-space of $\gradR$ becomes (at least) $n$ dimensional everywhere, with conserved directions $\{\vec{g}^i\}_{i \in \vec\phi}$ and tangent vectors $\{\vec{g}_i\}_{i \in \vec\phi}$ associated with each $\phi^i$.
Each $\vec{g}^i$ is a global constant and must have strictly non-negative entries to be physically meaningful, and $\phi^i = \vec{g}^i \vrho$.
The tangent space of the reactive nullcline is spanned by $\{\vec{g}_i\}_{i\in\vec\phi}$.
We can choose any convenient basis for the remaining $(m-n)$ non-conserved directions, such as a fixed set of orthogonal tangent vectors giving us a complete set $\{\vec{g}_1, \cdots, \vec{g}_m\}$.

For $n > 1$ we cannot necessarily satisfy the biorthogonality condition
  $\vec{g}^j \vec{g}_i = \delta_i^j
  =
  \vec{g}^i \vec{g}_j = \delta_j^i$
everywhere because this implies the existence of a global chart for the nullcline; this requires the nullcline to be fully integrable (in the sense of Frobenius theorem).
In general we can only construct a local coordinate chat in the local neighbourhood  $U \subseteq \chemicalspace$ surrounding some reference concentration $\widehat\vrho$.
We assume $U$ contains all the states relevant to the dynamics at long-times, which may require careful choice of the parameters $\vec\phi$ and $\widehat\vrho$.
The dual basis of one-forms $\{\vec{g}^1, \cdots, \vec{g}^m\}$ is then constructed within $U$ from the biorthogonality condition.

We introduce a set of coordinates $\vec\x = (\x^1, \cdots, \x^m)^\top$ associated with each tangent vector that charts $U$.
As \eqref{eq:differential-basis} does not necessarily hold globally with $n > 1$ we must proceed with \eg perturbation expansions of \eqref{eq:rho-contour} around $\widehat\vrho$.
We can expand the tangent vectors about $\widehat\vrho$ as
\begin{equation*}
  \vec{g}_i
  =
  \hat{\vec{g}}_i
  + \frac{\x^j}{1!} \hat{\vec{g}}_{i,j}
  + \frac{\x^j \x^k}{2!} \hat{\vec{g}}_{i,jk}
  + \cdots\,.
\end{equation*}
Recall that the hats $\hat{(\cdot)}$ indicate a quantity is evaluated at $\widehat\vrho$.
A good coordinate basis is \emph{torsion-free}, meaning $\vec{g}_{i,j} = \vec{g}_{j,i}$ and the contour \eqref{eq:rho-contour} becomes path-independent.
In this case we obtain the local expansion
\begin{equation*}
  \vrho
  =
  \widehat\vrho
  + \frac{\x^i}{1!} \hat{\vec{g}}_i
  + \frac{\x^i \x^j}{2!} \hat{\vec{g}}_{i,j}
  + \frac{\x^i \x^j \x^k}{3!} \hat{\vec{g}}_{i,jk}
  + \cdots\,.
\end{equation*}
We can similarly obtain more general vector quantities such as the chemical flux $\vRflux$ by contour integration.
This contour becomes path-independent $\vRflux_{,ij} = \vRflux_{,ji}$ in a torsion-free basis because chemical space is (conformally) flat.
We then obtain
\begin{equation*}
  \vRflux
  =
  \refvRflux
  + \frac{\x^j}{1!} \refvRflux_{,j}
  + \frac{\x^j \x^k}{2!} \refvRflux_{,jk}
  + \frac{\x^j \x^k \x^l}{3!} \refvRflux_{,jkl}
  + \cdots\,.
\end{equation*}

\subsection{Limiting field theory}
\label{sec:vector-bulk}

The procedure to obtain the limiting field theory is identical to \secref{sec:interface-dynamics}.
Following adiabatic elimination of $\vrho_\ex$, the first derivatives of $\vrho$ must now have a sum over thes components of $\vec\phi$, \ie
\begin{equation*}
  \partial_t \vrho = \left(\partial_t \phi^i\right) \vec{g}_i
  \quad \textrm{and} \quad
  \grad{\vrho} = \vec{g}_i \otimes \grad{\phi^i}\,.
\end{equation*}
The adiabatic eliminated solution $\vrho = \left( 1 + \LR^{-1} \LD \nabla^2 \right)^{-1} \vrho_\phi$ follows from \eqref{eq:adiabatic}, albeit the nullcline becomes $\vrho_\phi = \vrho_\phi(\vec\phi)$.
The leading term is simply $\vrho_\phi$, and the leading correction is the excess mode
\begin{equation*}
\begin{split}
  \vwrho
  &=
  -\LR^{-1} \LD \nabla^2 \vrho_\phi
  + \order{\nabla^4}
  \\ &=
  -\LR^{-1} \LD
  \sum_{i \in \vec\phi} \bigg(
    \vec{g}_i \nabla^2 \phi^i
    + \sum_{j \in \vec\phi}
    \vec{g}_{i,j} \nabla\phi^i \cdot \nabla\phi^j
  \bigg)
  + \order{\nabla^4}\,.
\end{split}
\end{equation*}
Inserting these into the evolution equation \eqref{eq:evolution} and projecting we obtain the \emph{vector} field theory
\begin{equation*}\label{eq:bulk-vector-theory}
  \partial_t \phi^i
  =
  \nabla^2{\left( \mu_\phi^i + \mu_\ex^i \right)}
  \qquad \forall \; i \in \{\phi^1, \cdots, \phi^n\}\,,
\end{equation*}
in terms of the bulk and excess effective chemical potentials $\mu_\phi^i \equiv \vec{g}^i \LD \vrho_\phi$ and $\mu_\ex^i \equiv \vec{g}^i \LD \vrho_\ex$.

The bulk free energy functional implied by $\vec{g}^i \LD \vrho_\phi$ again emerges simply from the shape of the nullcline (which is now $n$-dimensional).
The usual techniques to determine bulk phase behaviour can be similarly applied to this vector theory.
Linear instability, for example, occurs when the mixed tensor $D_j^i \equiv \vec{g}^i \LD \vec{g}_j$ develops a negative eigenvalue in this basis.
Note that $D_j^i$ are simply the coefficients of $\LD$ decomposed into our chemical basis, \ie $\LD = D_j^i \, \vec{g}_i \otimes \vec{g}^j$.
These components are effective transport coefficients, which will vary throughout chemical space as $\LD$ is parallel transported.
The constancy of $\LD$ is encoded in a vanishing \emph{covariant} derivative.

The interface correction to the chemical potential is then obtained similarly from inserting the expansion of $\vwrho$ into $\wmu = \vec{g}^\phi \LD \vwrho$.
Separate integrable and non-integrable interface terms can be pulled out of $\wmu$ in much the same way as we determined for the scalar theory, but the specifics of this are beyond the scope of the present work.

\section{Canonical linear stability analysis}
\label{appendix:linear-stability}

In the main text we worked out a fully nonlinear theory.
Here we verify its consistency with the result of the more conventional linear stability analysis about some homogeneous reference state $\widehat\vrho$.
In Fourier space \eqref{eq:evolution} becomes
\begin{equation}
  \partial_t \widetilde{\vrho}
  =
  \vec{L} \widetilde{\vrho}
  + \order{|\vrho|^2}\,,
\end{equation}
where a tilde $\widetilde{(\cdot)}$ indicates the Fourier transform and introducing the linear operator $\vec{L} = \refgradR - q^2 \LD$.
Note that $\vec{L} \to \LR = \chemicalgrad \vRflux$ as $q \to 0^+$.
This becomes linearly unstable to inhomogeneous fluctuations when $\vec{L}$ develops a positive eigenvalue.
We write the $q$-dependent eigenvalues of $\vec{L}$ as $\{\eval_i(q)\}$.
Writing the eigenvalues of $\refgradR$ at some reference state as $\{\hat\eval_i\}$, we know that $\eval_i(q) \to \hat\eval_i$ as $q \to 0$.

Projecting the $i$th eigenvalue problem $\vec{L} \, \widetilde{\vec{g}}_i = \eval_i \vec{g}_i$ gives the following expression for the wavevector-dependent eigenvalue
\begin{equation*}
  \eval_i(q)
  =
  \vec{g}^i(q) \left( \refgradR - q^2 \LD \right) \vec{g}_i(q)\,.
\end{equation*}
We look for the expansions
\begin{align*}
  \eval_i(q) &= \eval_i^{(0)} + q^2 \eval_i^{(2)} + q^4 \eval_i^{(4)} + \order{q^6}\,,
  \\
  \vec{g}_i(q) &= \vec{g}_i^{(0)} + q^2 \vec{g}_i^{(2)} + q^4 \vec{g}_i^{(4)} + \order{q^6}\,,
  \\
  \vec{g}^i(q) &= \vec{g}^i_{(0)} + q^2 \vec{g}^i_{(2)} + q^4 \vec{g}^i_{(4)} + \order{q^6}\,,
\end{align*}
with the $q = 0$ contributions corresponding to the homogeneous reference state \ie $\eval_i^{(0)} = \hat\eval_i$, $\vec{g}_i^{(0)} = \hat{\vec{g}}_i$ and $\vec{g}^i_{(0)} = \hat{\vec{g}}^i$.
Inserting these expansions into the eigenvalue problem gives corrections up to $\order{q}^4$
\begin{align*}
  \eval_i^{(2)}
  &=
  - \vec{g}_{(0)}^i \LD \vec{g}_i^{(0)}
  =
  - \hat{D}_i^i
  \\
  \eval_i^{(4)}
  &=
  \vec{g}_{(2)}^i \refgradR \vec{g}_i^{(2)}
  - \vec{g}_{(2)}^i \LD \vec{g}_i^{(0)}
  - \vec{g}_{(0)}^i \LD \vec{g}_i^{(2)}\,.
\end{align*}
In order to eliminate terms like
\begin{equation*}
  \vec{g}_{(0)}^i \refgradR \vec{g}_i^{(2)}
  =
  \hat\eval_i \, \vec{g}_{(0)}^i \vec{g}_i^{(2)}
\end{equation*}
we had to make use of the orthogonality of $\vec{g}_i^{(0)}$ with its higher-order corrections.

For $\eval_i^{(4)}$ we need the leading correction to the eigenvectors.
Eigenvector perturbation theory gives these as
\begin{align*}
  \vec{g}_i^{(2)}
  &=
  \sum_{j \ne i}
  \frac{\vec{g}_{(0)}^j \LD \vec{g}_i^{(0)}}{\hat\eval_j - \hat\eval_i}
  \vec{g}_j^{(0)}
  =
  \sum_{j \ne i}
  \frac{\hat{D}_i^j}{\hat\eval_j - \hat\eval_i}
  \vec{g}_j^{(0)}\,,
  \\
  \vec{g}_{(2)}^i
  &=
  \sum_{j \ne i}
  \frac{\vec{g}_{(0)}^i \LD \vec{g}_j^{(0)}}{\hat\eval_j - \hat\eval_i}
  \vec{g}_{(0)}^j
  =
  \sum_{j \ne i}
  \frac{\hat{D}_j^i}{\hat\eval_j - \hat\eval_i}
  \vec{g}_{(0)}^j\,.
\end{align*}
Inserted into the $\order{q^4}$ correction to the eigenvalues, we find
\begin{equation*}
  \eval_i^{(4)}
  =
  \sum_{j \ne i}
  \frac{\hat{D}_j^i \hat{D}_i^j}{\hat\eval_j - \hat\eval_i}
  \left(
  \frac{\hat\eval_j}{\hat\eval_j - \hat\eval_i} - 2
  \right)\,.
\end{equation*}
Along a conserved (zero) eigenvalue, this becomes simply
\begin{equation}
  \eval_i^{(4)}
  =
  - \sum_{j \ne i}
  \frac{\hat{D}_j^i \hat{D}_i^j}{\hat\eval_j}
  \equiv
  - \kappa_0\,.
\end{equation}
As we have assumed the eigenvalues are unique, this expression is only valid when there is a single conservation law in the $i$th direction.

For instabilities along a conserved direction, we find an evolution of the normal mode
\begin{equation*}
  \partial_t \widetilde\phi = \epsilon_\finescale q^2 - \kappa_0 q^4 + \order{q^6}\,,
\end{equation*}
which for unstable solutions $\epsilon_\finescale > 0$ we find the most unstable (maximum) wavenumber
\begin{equation*}
  q^*
  =
  \sqrt{\frac{\epsilon_\finescale}{2\kappa}}
  =
  \left(
  \sum_{j\ne\phi} \frac{2 \hat{D}_j^\phi \hat{D}_\phi^j}{\hat\eval_j \hat{D}_\phi^\phi}
  \right)^{-1/2}
\end{equation*}
where we use $\phi$ also as label for the index of the direction along the conserved order parameter $\phi = \vec{g}^\phi \vrho$.
Back in real space, the order parameter evolves as
\begin{equation*}
  \partial_t \phi
  =
  \nabla^2{\left(
    - \epsilon_\finescale \phi - \kappa_0 \nabla^2 \phi
  \right)}
  + \order{\phi^2, \nabla^6}\,,
\end{equation*}
where $\epsilon_\finescale = -\hat{D}_\phi^\phi = -\vec{g}^\phi \LD \vec{g}_\phi$ drives the instability and $\kappa_0$ plays the role of an effective surface-tension.
This agrees with the linear terms in the nonlinear theory we derived in the main text.
In particular, we obtained $\epsilon_\finescale > 0$ as the spinodal condition in \eqref{eq:instability-parameter} and $\kappa_0$ is the leading term in the expansion of $\kappa(\phi) = \vec{g}^\phi \LD \LR^{-1} \LD \vec{g}_\phi$ from \eqref{eq:kappa-lambda}.

\section{Negative cross-diffusion is needed for Turing instabilities in ideal systems with monotonic reactive nullclines}
\label{appendix:cross-diffusion}

In order to acquire a negative $\eval_i^{(2)}(q)$ along the $i$th conserved direction we must have $\hat{D}_i^i \equiv \hat{\vec{g}}^i \LD \hat{\vec{g}}_i < 0$.
We drop the hats from quantities below, so it is implied that we are evaluating the basis vectors at the reference state.

The object $D_i^i$ is a generalised inner product in the space of $\LD$: negative terms only arise when $\vec{g}^i$ and $\vec{g}_i$ have opposite signs along one of the principal axes of $\LD$.
This situation requires an eigenvector of $\LD$ to define a hyperplane (containing the origin) which partitions $\vec{g}^i$ and $\vec{g}_i$ on either side.
$\vec{g}^i$ defines the conservation law and so it must have non-negative entries, but $\vec{g}_i$ can have negative entries if the reactive nullcline is reentrant.

If the Cartesian entries of $\LD$ are all positive, the Perron-Frobenius theorem guarantees that the dominant eigenvector lies in the physical orthant.
In this case a simple way to achieve $D_i^i < 0$ is if $\vec{g}_i$ has negative entries.
This occurs at the inflection points in the ``N-type'' reactive nullclines sketched in \Figref{fig:phase-portraits}(b-c).

Otherwise, if $\vec{g}_i$ lies in the physical orthant as in \Figref{fig:phase-portraits}(a) then negative off-diagonal entries of $\LD$ are needed.
The diagonal entries of $\LD$ are positive since it is symmetric and positive-definite.
We write $\LD = \LD_\diag + \LD_\cross$, where $\LD_\diag$ is the diagonal matrix with the same diagonal entries as $\LD$, and $\LD_\cross$ is the matrix of off-diagonal entries.
Then
\begin{equation*}
  D_i^i
  =
  \vec{g}^i \LD_\diag \vec{g}_i
  + \vec{g}^i \LD_\cross \vec{g}_i\,.
\end{equation*}
Then, instability occurs when
\begin{equation*}
  \vec{g}^i \LD_\cross \vec{g}_i
  <
  -|\vec{g}^i \LD_\diag \vec{g}_i|\,.
\end{equation*}
The right-hand side is negative since $\vec{g}^i$, $\vec{g}_i$ and $\LD_\diag$ each have strictly non-negative entries.
This condition can only be met if $\LD_\cross$ has negative entries, demonstrating that $D_i^i < 0$ requires $\LD$ to have negative cross-diffusion terms.

For systems lacking explicit forces we have shown that linear instability towards an inhomogeneous state only occurs as $q \to 0^+$ when either there is an inflection point in the reactive nullcline, or there are negative phoretic transport coefficients.
Phoretic terms should really emerge from forces, so the latter situation is not properly consistent.
However, this provides a bridge to systems with forces where $\LD \to \LD\vec{B}$ where $\vec{B}$ is the matrix of second-virial coefficients; repeating the analysis above suggests that attractions can drive instabilities for systems with nullclines in the form of \Figref{fig:phase-portraits}(a).

\section{Asymptotic solution to adiabatic elimination in the linear operator limit is consistent with the Green's function solution}
\label{appendix:green}

In this note we aim to bolster confidence in the adiabatically eliminated solution we derived in the main text (\eg in \secref{sec:simple-adiabatic}) through asymptotic arguments, \ie
\begin{align}
  \label{eq:screened-poisson-expansion}
  \vwrho(\vec{r})
  &=
  (\Id + \LR^{-1} \LD \nabla^2)^{-1} \vec{u}(\vec{r})
  \\ \nonumber &=
  \vec{u}(\vec{r})
  - \LR^{-1} \LD \nabla^2 \vec{u}(\vec{r})
  + (\LR^{-1} \LD \nabla^2)^2 \vec{u}(\vec{r})
  - \cdots
\end{align}
where $\vec{u}(\vec{r}) = -\LR^{-1} \LD \nabla^2 \vrho_\phi$.
The central claim of \eqref{eq:screened-poisson-expansion} is that the operator inverse of $(\Id + \LR^{-1} \LD \nabla^2)^{-1}$ can be expanded in powers of \emph{local} derivatives of $u(\vec{r})$.
As operator inverses normally involve integrals (via Green's function solutions), this result is slightly surprising as it prohibits truly non-local behaviour.
In principle, patterns could be intrinsically non-local, so it is worth exploring where our local series expansion \eqref{eq:screened-poisson-expansion} is consistent with the non-local Green's function solution in the linear operator limit where $\LR^{-1}$ is a constant.

To simplify calculation, we consider the scalar limit where $\rho_\ex, u \in \mathbb{R}$; the extension to vector quantities is straightforward.
We write $\LR^{-1} \LD = \pm \ell^2$ where $\ell$ has units of length, with $-\ell^2$ where $\LR$ is stable and $+\ell^2$ in an unstable region (\eg on the unstable branch in a reentrant nullcline).
We consider the latter for the analytic continuation of the theory into the unstable region, but note that this entire approach breaks down at the turning points separating stable annd unstable regions where $\ell \to \infty$.
The linearised form of \eqref{eq:screened-poisson-expansion} then becomes
\begin{equation}\label{eq:screened-poisson-expansion-linear}
  \begin{split}
    \rho_\ex(\vec{r})
    &=
    (1 - \ell^2 \nabla^2)^{-1} u(\vec{r})
    \\ &=
    u(\vec{r}) + \ell^2 \nabla^2 u(\vec{r}) + \ell^4 \nabla^4 u(\vec{r}) + \cdots
  \end{split}
\end{equation}
\ie with $n$th term is $\ell^{2n} \nabla^{2n} u(\vec{r})$.

To verify consistency, we must show \eqref{eq:screened-poisson-expansion-linear} is equivalent to the Green's function solution
\begin{equation*}
  \rho_\ex(\vec{r}) = \int \dd\vec{r}' \mathcal{G}(\vec{r}, \vec{r}') u(\vec{r}')\,.
\end{equation*}
The Green's function is the solution to the screened Poisson equation
\begin{equation*}
  (1 - \ell^2 \nabla^2) \mathcal{G} = - \delta(\vec{r} - \vec{r}')
\end{equation*}
The solution can then be found via Fourier transformation and contour integrating (or otherwise), giving
\begin{equation}\label{eq:screened-poisson-G}
  \mathcal{G}(r = |\vec{r} - \vec{r}'|)
  =
  \frac{K_{\frac{d}{2}-1}{\left( \frac{r}{\ell} \right)}}{2\pi \ell^2 (2\pi r\ell)^{\frac{d}{2}-1}}\,,
\end{equation}
for $d \ge 2$, and where $\{K_n\}$ are the modified Bessel functions of the second kind.
In physical dimensions it is easily verified that this reduces to the known results
\begin{equation*}
  \mathcal{G}(r)
  =
  \begin{cases}
    \frac{K_0{\left( \frac{r}{\ell} \right)}}{2\pi \ell^2} & d = 2 \\
    \frac{\exp{\left(-\frac{r}{\ell}\right)}}{4\pi \ell^2 r} & d = 3\,.
  \end{cases}
\end{equation*}

In order to show equivalence, we expand the inhomogeneous term as
\begin{equation*}
  u(\vec{r}')
  =
  u(\vec{r})
  + \nabla u(\vec{r}) \cdot \left(\vec{r}' - \vec{r}\right)
  + \frac{1}{2!} \nabla \nabla u(\vec{r}) : \left(\vec{r}' - \vec{r}\right)^{\otimes 2}
  + \cdots\,,
\end{equation*}
and then insert this into the Green's function solution.
The inhomogeneous solution then becomes
\begin{equation*}
  \begin{split}
    \rho_\ex(\vec{r})
    =
    &\; u(\vec{r}) \int \dd\vec{r}' \, \mathcal{G}(r')
    \\ &+ \nabla u(\vec{r}) \cdot \int \dd\vec{r}' \, \mathcal{G}(r') \vec{r}'
    \\ &+ \frac{\nabla \nabla u(\vec{r})}{2!} :
    \int \dd\vec{r}' \, \mathcal{G}(r') \vec{r}' \otimes \vec{r}'
    + \cdots
  \end{split}
\end{equation*}
As $\vec{r}'$ has odd parity whereas $\mathcal{G}(|\vec{r}'|)$ has even parity, only integrals involving even numbers of $\vec{r}'$ will survive integration.
Only $\nabla^{\otimes 2n} u$ therefore contribute to $\vwrho$.
The leading term is simply proportional to $u(\vec{r})$.
The next leading term is then
\begin{equation*}
  \frac{\nabla\nabla u}{2!} : \int \dd\vec{r}' \, \mathcal{G}(r') \vec{r}' \otimes \vec{r}'
  =
  \frac{1}{2!} \sum_{i = j = 1}^d \frac{\partial^2 u(\vec{r})}{\partial x_i \partial x_j}
  \int \dd\vec{r}' \, \mathcal{G}(r') x_i' x_j'
\end{equation*}
where $\{x_1', \cdots, x_d'\}$ are the Cartesian components of $\vec{r}'$.
To proceed, we focus on the inner product with $\vec{e}_i \otimes \vec{e}_j$.
This integral must vanish when $i \ne j$ because the Cartesian components have odd parity, but their square becomes even when $i = j$.
The integral for each diagonal entry must be the same by rotation invariance, so that the final result involves the trace $\Tr{(\nabla\nabla u)} = \nabla^2 u$ times a radial integral over $G(r')$.
So far this is consistent with our asymptotic solution, but we need the constant of proportionality (which should be $\ell^{2n}$) and we need to extend this to the $n$th term.

Continuing to higher-order terms is more intricate.
First, we note that the $n$th harmonic operator gives
\begin{equation*}
  \nabla^{2n} u(\vec{r})
  =
  \sum_{|\alpha| = n}
  \frac{\partial^{2n} u(\vec{r})}{\partial x_1^{2\alpha_1} \cdots \partial x_d^{2\alpha_d}}\,.
\end{equation*}
The $n$th non-zero term in our expansion involves a sum over $|\beta| = 2n$, but only terms involving odd derivatives survive (by an identical argument as above), so we can transform the sum to one over $|\alpha| = n$:
\begin{equation*}
  \sum_{\substack{|\beta| = 2n}}
  \frac{\partial^{2n}}{\partial x_{\beta_1} \cdots \partial x_{\beta_{2n}}}
  \to
  \sum_{\substack{|\alpha| = n}}
  \frac{\binom{2n}{2\alpha}}{\binom{n}{\alpha}}
  \frac{\partial^{2n}}{\partial x_1^{2\alpha_1} \cdots \partial x_d^{2\alpha_d}}\,,
\end{equation*}
where the multinomial coefficients evaluate to
\begin{equation*}
  \frac{\binom{2n}{2\alpha}}{\binom{n}{\alpha}}
  =
  \frac{(2n)!}{(2\alpha_1)! \cdots (2\alpha_d)!} \frac{\alpha_1! \cdots \alpha_d!}{n!}\,.
\end{equation*}
The remaining contribution in the $n$th term is an the integral over $2n$ Cartesian coordinates (and a combinatorial prefactor from the Taylor expansion):
\begin{equation*}
  \frac{1}{(2n)!}
  \int \dd\vec{r} \, \mathcal{G}(r) x_1^{2\alpha_1} \cdots x_d^{2\alpha_d}
  =
  \frac{\Theta_n}{(2n)!} \int \dd r \, r^{2n + d - 1} \mathcal{G}(r)\,.
\end{equation*}
On the right-hand side we decomposed it into an angular integral $\Theta_n$ and a radial integral.
This step makes use of hyperspherical coordinates with radius $r$ and angular coordinates $\{\theta_1, \cdots, \theta_{d-1}\}$ (with $\theta_1 \in [0, 2\pi]$ and $\theta_i \in [0, \pi]$ for $i > 1$).
The angular integral measures the distribution of the monomial $(x_1^{\alpha_1} \cdots  x_d^{\alpha_d})$ on the unit $(d-1)$-sphere.
Note that this monomial has degree $2|\alpha| = 2n$.
Given explicitly, the angular integral is
\begin{equation*}
  \Theta_n
  =
  \int_{S^{d-1}} \dd\Omega_d \, x_1^{\alpha_1} \cdots  x_d^{\alpha_d}
\end{equation*}
where $S^n$ is the unit $n$-sphere.
Note that inside this integral $|\vec{x}| = 1$ as we are constrained to the unit sphere.
We employ the trick of considering the related integral \cite{folland2001}
\begin{equation*}
  \begin{split}
    \int_{\mathbb{R}^d} \dd \vec{r} \, e^{-r^2} x_1^{2\alpha_1} \cdots x_d^{2\alpha_d}
    &=
    \Theta_n \int_0^\infty \dd r \, r^{2n + d-1} e^{-r^2}
    \\ &=
    \frac{\Theta_n}{2} \Gamma{\left( \frac{2n + d}{2} \right)}
  \end{split}
\end{equation*}
But the left-hand side can also be evaluated in Cartesian coordinates:
\begin{equation*}
  \begin{split}
  &\int_{\mathbb{R}^d} \dd \vec{r} \, e^{-r^2} x_1^{2\alpha_1} \cdots x_d^{2\alpha_d}
  =
  \prod_{i=1}^d \left( \int_{-\infty}^\infty \dd x_i \, x_i^{2 \alpha_i} e^{-x_i^2} \right)\,,
  \\ =
  &\prod_{i=1}^d \Gamma{\left( \alpha_i + \frac{1}{2} \right)}
  =
  \prod_{i=1}^d \frac{\sqrt{\pi} (2 \alpha_i)!}{4^{\alpha_i} \alpha_i!}
  =
  \frac{\pi^{\frac{d}{2}}}{4^n} \prod_{i=1}^d \frac{(2 \alpha_i)!}{\alpha_i!}\,.
  \end{split}
\end{equation*}
Equating these two approaches, we find
\begin{equation*}
  \Theta_n
  =
  \frac{
    2 \pi^{\frac{d}{2}}
  }{
    4^n \Gamma{\left( \frac{2n + d}{2} \right)}
  }
  \prod_{i=1}^d \frac{(2 \alpha_i)!}{\alpha_i!}\,.
\end{equation*}
Returning to our expansion, we have $n$th non-zero term
\begin{equation*}
  \begin{split}
  \frac{1}{(2n)!}
  \sum_{\substack{|\beta| = 2n}}
  \frac{\partial^{2n} u(\vec{r})}{\partial x_{\beta_1} \cdots \partial x_{\beta_{2n}}}
  \int \dd\vec{r}' \mathcal{G}(r') x_1^{\alpha_1} \cdots x_d^{\alpha_d}
  \\ =
  \left(
  \frac{
    2 \pi^{\frac{d}{2}}
  }{
    n! 4^n \Gamma{\left( \frac{2n + d}{2} \right)}
  }
  \int \dd r' \, (r')^{2n + d - 1} \mathcal{G}(r')
  \right)
  \nabla^{2n} u(\vec{r})\,.
  \end{split}
\end{equation*}
This confirms that the $n$th term is indeed proportional to $\nabla^{2n} u$, as we found for the leading correction.
To obtain the proportionality constant we insert the Green's function \eqref{eq:screened-poisson-G} into the integral, which then evaluates to
\begin{equation*}
  \begin{split}
    \frac{
      (2 \ell)^{1 - \frac{d}{2}}
    }{
      n! 4^n \ell^2 \Gamma{\left( \frac{2n + d}{2} \right)}
    }
    \int_0^\infty \dd r \, r^{2n + \frac{d}{2}}
    K_{\frac{d}{2}-1}{\left( \frac{r}{\ell} \right)}
    =
    \ell^{2n}\,.
  \end{split}
\end{equation*}
This is identical to the result of our asymptotic analysis \eqref{eq:screened-poisson-expansion-linear}, and so we have demonstrated equivalence of the approaches (at least for constant $\LR$, and in the one-component case).

In \secref{sec:worked-example} we analytically continue our results inside a region where the nullcline is reentrant.
In that region the adiabatic equation is a Helmholtz equation instead of a screened Poisson equation.
To repeat the same steps for the Helmholtz equation, we make the change $\ell^2 \to - \ell^2$ in \eqref{eq:screened-poisson-expansion-linear} so that the expansion becomes
\begin{equation}\label{eq:helmholtz-expansion-linear}
  \begin{split}
    \rho_\ex(\vec{r})
    &=
    (1 + \ell^2 \nabla^2)^{-1} u(\vec{r})
    \\ &=
    u(\vec{r}) - \ell^2 \nabla^2 u(\vec{r}) + \ell^4 \nabla^4 u(\vec{r}) - \cdots\,.
  \end{split}
\end{equation}
The steps to find the inverse via Green's functions would be identical to before, except in the penultimate line where we evaluate the radial integral we would make use of the Green's function for the Helmholtz equation.
The resulting integral is more tricky (as $G(r)$ is oscillatory), but as we have increased confidence in the expansion solution we can afford to be more specialised.
For $d = 3$, the Green's function is the familiar spherical wave kernel
\begin{equation*}
  \mathcal{G}(r = |\vec{r} - \vec{r}'|)
  =
  \frac{\exp{\left( \frac{i r}{\ell} \right)}}{4 \pi r}\,.
\end{equation*}
For $d = 3$ we evaluate the integral for the $n$th term as
\begin{equation*}
  \begin{split}
    \lim_{\epsilon \to 0^+}
    \frac{
      2\pi^{\frac{3}{2}}
    }{
      n! 4^n \Gamma{\left( \frac{2n + 3}{2} \right)}
    }
    \int_0^\infty \dd r \, r^{2n + 1} e^{i \frac{r}{\ell} - \epsilon r}
    =
    -(-1)^n \ell^{2n}
  \end{split}
\end{equation*}
giving \eqref{eq:helmholtz-expansion-linear}.

Together \eqref{eq:screened-poisson-expansion-linear} (in arbitrary $d$) and \eqref{eq:helmholtz-expansion-linear} (in $d = 3$) confirm \eqref{eq:screened-poisson-expansion} for scalar quantities and in the limit where the $\LR^{-1} \sim \ell^2$ is a constant.
The expansions we use in the main text therefore seem well-justified where $|\ell|^2 \ll 1$.
For larger $|\ell|$ we would expect the series solution \eqref{eq:screened-poisson-expansion-linear} to not converge, in which case the exact Green's function solution is genuinely non-local.
In Fourier space, the series solution can be written has
\begin{equation*}
  \tilde\rho_\ex(q) = \frac{\tilde{u}(q)}{1 + \ell^2 q^2}
\end{equation*}
which diverges as $|\ell q|^2 \to 1$.
This threshold indicates where the chemical flux becomes too weak to confine the system in the vicinity of the nullcline.
We note that $\ell \to \pm \infty$ at turning points demarcating stable and unstable regions.
This reconfirms that reentrant nullclines are pathological to the asymptotic series solution developed here.

\section{Beyond onset: adiabatic elimination with an arbitrary reference contour}
\label{appendix:arbitrary-reference}

Strictly speaking, our analysis in \secref{sec:interface-dynamics} is only applicable around the onset of phase separation where the homogeneous solution (along the nullcline) remains a good reference state.
Here we show that our theory generalises to more arbitrary reference contours, demonstrating that the resulting field theory \eqref{eq:active-model-b} remains valid beyond onset and even approaching the steady-state.

We introduce a new point-wise reference geometry $\bar\vrho_\phi = \bar\vrho_\phi(\phi)$ which can be written as the contour integral
\begin{equation}
  \bar\vrho_\phi(\phi)
  =
  \widehat\vrho
  + \int_0^\phi \dd\varphi \, \bar{\vec{g}}_\phi(\varphi)
\end{equation}
similarly to \eqref{eq:scalar-nullcline} but with a different tangent vector $\bar{\vec{g}}_\phi$.
However, now the flux does not necessarily vanish along this contour.
The expansion of the flux then becomes
\begin{equation*}
  \vRflux = \bar\vRflux(\phi) + \bar\LR(\phi) (\vrho - \vrho_\phi) + \order{\vrho^2}\,,
\end{equation*}
where $\bar\vRflux \equiv \vRflux(\vrho = \bar\vrho_\phi)$ and $\bar\LR \equiv (\chemicalgrad \vRflux)_{\vrho=\bar\vrho_\phi}$ are now evaluated at points along the new contour.
Writing $\bar\vrho_\ex = \vrho - \bar\vrho_\phi$, then repeating the steps of \secref{sec:simple-adiabatic} would result in solution
\begin{subequations}\label{eq:arbitrary-reference}
\begin{align}
\bar\vrho_\ex
&=
-\left(\bar\LR + \LD \nabla^2 \right)^{-1}
\left( \bar\vRflux + \LD \nabla^2 \bar\vrho_\phi \right)\,,
\\ \nonumber &=
\sum_{n=1}^\infty \left( -\bar\LR^{-1}(\phi) \LD \nabla^2 \right)^n
\left( \bar\vrho_\phi(\phi) - \bar\LR^{-1}(\phi) \bar\vRflux(\phi) \right)\,,
\end{align}
or equivalently the total concentration is
\begin{align}\label{eq:adiabatic-solution-arbitrary-contour}
\vrho
&=
\left(\Id + \bar\LR^{-1} \LD \nabla^2 \right)^{-1}
\left( \bar\vrho_\phi - \bar\LR^{-1} \bar\vRflux \right)\,,
\\ \nonumber &=
\sum_{n=0}^\infty \left( -\bar\LR^{-1} \LD \nabla^2 \right)^n
\left( \bar\vrho_\phi - \bar\LR^{-1} \bar\vRflux \right)\,.
\end{align}
\end{subequations}
Inserting this back into \eqref{eq:evolution} and projecting onto the order parameter results in a field theory identical in form to that previously obtained in \eqref{eq:active-model-b}, albeit with different coefficients.
It is straightforward to show by substitution that if the flux is exactly linear around the nullcline \ie $\bar\vRflux = \bar\LR (\bar\vrho_\phi - \vrho_\phi)$ (recalling that $\vrho_\phi$ is our original nullcline reference geometry), then \eqref{eq:arbitrary-reference} reduces to \eqref{eq:adiabatic-solution} as expected.

For a sufficiently good guess for the contour $\bar\vrho_\phi = \bar\vrho_\phi(\phi)$, the resulting coefficients would be more accurate at longer times when deviations from the nullcline become pronounced.
The strength of the theory approaching the steady-state then depends on finding accurate approximations for the reference contour $\bar\vrho_\phi$.
We note that for two-component $m = 2$ systems the steady-state must fall along a straight line with $\vec{g}^\phi \LD \vrho = \mathrm{const}$, and so it is trivial to construct a theory that becomes \emph{exact} approaching the steady-state.
For more components a reasonable guess would be the steepest descent path along the plane normal to $\vec{g}^\phi \LD$.

\section{Weakly nonlinear analysis}
\label{appendix:bergmann}

Here we compare our approach with the weakly nonlinear analysis of \citeauthor{bergmann2018} \cite{bergmann2018}.
We assume the system has just reached the linear instability threshold, from some initial homogeneous configuration $\widehat\vrho$ with zero flux $\vRflux(\vrho = \widehat\vrho) = 0$.
The instability parameter $\epsilon_\finescale = -\vec{g}^\phi \LD \widehat{\vec{g}}_\phi$ (determined from the linear stability analysis in Appendix~\ref{appendix:linear-stability}) is small by definition at threshold.
The perturbation $\delta \vrho \equiv \vrho - \widehat\vrho$ then evolves as
\begin{equation*}
\partial_t \delta\vrho = \left( \widehat\LR + \LD \nabla^2 \right) \delta\vrho
+ \vec{N}
\end{equation*}
where $\widehat{\LR}$ is evaluated at $\widehat\vrho$ (so is genuinely linear now) and $\vec{N}$ are the nonlinear terms in $\vRflux$.
We then expand
\begin{equation*}
\delta\vrho = \sum_{n=1}^\infty \epsilon_\finescale^{n/2}\delta\vrho^{(n)}, \qquad \vec N = \sum_{n=2}^\infty\epsilon_\finescale^{n/2}\vec N^{(n)}
\end{equation*}
together with rescaling time and length through
\begin{equation*}
  \nabla\to\epsilon_\finescale^{1/2}\nabla, \qquad \partial_t \to \epsilon_\finescale^{3/2}\partial_{T_3} + \epsilon_\finescale^2\partial_{T_4}\,.
\end{equation*}
These particular scalings can be determined as one of the simplest valid choices remaining after carefully eliminating physically implausible dominant balances.
The evolution equation now reads
\begin{equation*}
  \epsilon_\finescale^{3/2}\partial_{T_3}\delta\vrho + \epsilon_\finescale^2\partial_{T_4}\delta\vrho = \widehat{\LR} \delta\vrho + \epsilon_\finescale \LD \nabla^2\delta\vrho + \vec N\,.
\end{equation*}
Inserting the expansions and (naively) collecting terms in powers of $\epsilon_\finescale$ leads to
\begin{align*}
  \epsilon_\finescale^{1/2}: \,
  \widehat{\LR} \delta\vrho^{(1)} &= 0\,, \\
  \epsilon_\finescale^{\hphantom{1/2}}: \,
  \widehat{\LR} \delta\vrho^{(2)} &= -\vec N^{(2)}\,, \\
  \epsilon_\finescale^{3/2}: \,
  \widehat{\LR} \delta\vrho^{(3)} &= -\vec N^{(3)} - \LD \nabla^2\delta\vrho^{(1)}\,, \\
  \epsilon_\finescale^{2\hphantom{/2}}: \,
  \widehat{\LR} \delta\vrho^{(4)} &= -\vec N^{(4)} - \LD \nabla^2\delta\vrho^{(2)} + \partial_{T_3}\delta\vrho^{(1)}\,, \\
  \epsilon_\finescale^{5/2}: \,
  \widehat{\LR} \delta\vrho^{(5)} &= -\vec N^{(5)} - \LD \nabla^2\delta\vrho^{(3)} + \partial_{T_3}\delta\vrho^{(2)} + \partial_{T_4}\delta\vrho^{(1)}\,.
\end{align*}
These equations are almost equivalent to those used by \citeauthor{bergmann2018} in deriving their solution, but the stated magnitudes in powers of $\epsilon_\finescale$ are misleading as we will demonstrate below.

Noting that the left-hand side always permits a solution in the right null-space of $\widehat{\LR}$, \ie proportional to $\widehat{\vec{g}}_\phi$, we decompose $\delta \vrho^{(i)} = \delta \phi^{(i)} \widehat{\vec{g}}_\phi + \delta\vrho_\ex^{(i)}$ with $\vec{g}^\phi \delta\rho_\ex = 0$.
The hierarchy above is conventionally solved by Fredholm alternative.
Contracting onto the left null space of $\widehat{\LR}$ (the conserved direction $\vec{g}^\phi$) gives the ostensible solvability criteria
\begin{align*}
0 &= \vec{g}^\phi \LD \nabla^2\delta\vrho^{(1)}\,, \\
\partial_{T_3} \delta \phi^{(1)} &= \vec{g}^\phi \LD \nabla^2\delta\vrho^{(2)}\,, \\
\partial_{T_3} \delta \phi^{(2)} + \partial_{T_4} \delta \phi^{(1)} &= \vec{g}^\phi \LD \nabla^2\delta\vrho^{(3)}\,,
\end{align*}
using $\vec{g}^\phi \delta \vrho^{(i)} = \delta \phi^{(i)}$.
Inserting in the decomposition of $\delta\vrho^{(i)}$ onto the right-hand sides leads to
\begin{equation*}
\vec{g}^\phi \LD \nabla^2 \delta \vrho_\ex^{(i)} - \epsilon_\finescale \nabla^2 \phi^{(i)}
\end{equation*}
using the definition of the small parameter.
So we see that the contribution from the order parameter is $\order{\epsilon_\finescale}$ smaller and must be moved further down in the hierarchy.
The corrected solvability criteria are
\begin{subequations}\label{eq:bergmann-conserved}
\begin{align}
\hphantom{\partial_{T_3} \delta \phi^{(2)} + \partial_{T_4} \delta \phi^{(1)}}
0 &= \vec{g}^\phi \LD \nabla^2\delta\vrho_\ex^{(1)}\,, \\
\hphantom{\partial_{T_3} \delta \phi^{(2)} +}
\partial_{T_3} \delta \phi^{(1)} &= \vec{g}^\phi \LD \nabla^2\delta\vrho_\ex^{(2)}\,, \\
\partial_{T_3} \delta \phi^{(2)} + \partial_{T_4} \delta \phi^{(1)} &= \vec{g}^\phi \LD \nabla^2\delta\vrho_\ex^{(3)} - \nabla^2 \phi^{(1)}\,,
\end{align}
\end{subequations}
where terms appearing in the same equation now have the same magnitude.
While \citeauthor{bergmann2018} \cite{bergmann2018} do have an equivalent of the $-\nabla^2 \phi^{(1)}$ term appearing in their $\order{\epsilon_\finescale^{5/2}}$ equation, the way they derive it is misleading.
The meaning of $\epsilon_\finescale$ is essential to correctly order terms by magnitude.
Finally we note that these solvability criteria are simply the projections of the evolution equation onto the conserved space \eqref{eq:evolution-finite-reactions}.

The three solvability criteria provide an evolution equation for $\phi$ inside a time window where only $\delta \vrho^{(1)}$, $\delta\vrho^{(2)}$ and $\delta\vrho^{(3)}$ contribute.
These are obtained by solving
\begin{subequations}\label{eq:bergmann-nonconserved}
\begin{align}
  \order{\epsilon_\finescale^{1/2}}: \;
  \widehat{\LR} \delta\vrho^{(1)} &= 0\,, \\
  \order{\epsilon_\finescale^{\hphantom{1/2}}}: \;
  \widehat{\LR} \delta\vrho^{(2)} &= -\vec N^{(2)}\,, \\
  \order{\epsilon_\finescale^{3/2}}: \;
  \widehat{\LR} \delta\vrho^{(3)} &= -\vec N^{(3)} - \project_\ex \LD \nabla^2\delta\vrho^{(1)}\,.
\end{align}
\end{subequations}
Note the $\project_\ex$ appearing in the third equation after applying the solvability criterion solution; these equations are therefore the equivalent of solving the adiabatic equation in the reactive subspace \eqref{eq:adiabatic} that we pursued in the main text.
The choices of scalings of terms has selected a time window where the adiabatic solution becomes exact.

At this point one would follow through by solving the reactive equations \eqref{eq:bergmann-nonconserved} and substituting them into the solvability equations \eqref{eq:bergmann-conserved} to obtain an evolution equation in $\phi$.
This solution relies on the dominant balance $\widehat\LR \delta\vrho^{(2)} = -\vec N^{(2)}$ \ie that the quadratic term is non-zero.
When this holds the solution will match our general solution.
More specifically, the solution will reduce to our general solution \eqref{eq:adiabatic-solution-arbitrary-contour} using a straight-line reference contour going through the reference point $\widehat\vrho$ and parallel to the tangent vector $\widehat{\vec{g}}_\phi$.
An advantage of our geometric approach is that we do not require any particular dominant balance.
By contrast, the above analysis would need to be repeated with a different dominant balance (presumably $\widehat\LR \delta\vrho^{(2)} = -\vec N^{(3)}$ assuming the cubic term is non-zero) if the quadratic term were zero.

\bibliography{selected}

\end{document}